\documentclass[10pt,A4paper,amsmath,amssymb,aps,etoolbox,floatfix,
longbibliography,nofootinbib,noshowpacs,onecolumn,preprintnumbers,reprint,
showkeys,showpacs,superscript address]{revtex4}

\newcommand{\bea}{\begin{eqnarray}}
\newcommand{\beq}{\begin{equation}}

\newcommand{\eea}{\end{eqnarray}}
\newcommand{\eeq}{\end{equation}}

\usepackage{amsmath}
\usepackage{amsfonts}
\usepackage{amssymb}
\usepackage{bm}       
\usepackage{caption}
\usepackage{comment}
\usepackage[usenames,dvipsnames]{color}
\usepackage{dcolumn}  
\usepackage{epsfig}
\usepackage{float}
\usepackage{graphics}
\usepackage{graphicx} 
\usepackage{natbib}
\usepackage{hyperref}

\usepackage{capt-of} 
\usepackage{placeins} 
\usepackage{psfrag}
\usepackage{times}
\usepackage[varg]{txfonts}
\usepackage{xcolor}
\usepackage{xspace}

\begin{document}

\title{Questioning the $H_0$ tension via the look-back time}

\author{Salvatore Capozziello}

\affiliation{{\mbox Dipartimento di Fisica "E. Pancini", Universit\`{a} degli Studi di Napoli  "Federico II",}\\ 
{\mbox Complesso Universitario Monte S. Angelo, Via Cinthia 9 Edificio G, 80126 Napoli, Italy,}}

\affiliation{{\mbox Istituto Nazionale di Fisica Nucleare, Sezione di Napoli,}\\
{\mbox Complesso Universitario Monte S. Angelo, Via Cinthia 9 Edificio G, 80126 Napoli, Italy,}}

\affiliation{{\mbox Scuola Superiore Meridionale, Largo San Marcellino 10, 80138 Napoli, Italy.}}

\author{Giuseppe Sarracino}

\affiliation{{\mbox Dipartimento di Fisica "E. Pancini", Universit\`{a} degli Studi di Napoli  "Federico II",}\\ 
{\mbox Complesso Universitario Monte S. Angelo, Via Cinthia 9 Edificio G, 80126 Napoli, Italy,}}

\affiliation{{\mbox Istituto Nazionale di Fisica Nucleare, Sezione di Napoli,}\\
{\mbox Complesso Universitario Monte S. Angelo, Via Cinthia 9 Edificio G, 80126 Napoli, Italy,}}

\author{Alessandro D.A.M. Spallicci}

\affiliation{\mbox{Laboratoire de Physique et Chimie de l'Environnement et de l'Espace, UMR 7328,}\\
{\mbox Centre National de la Recherche Scientifique,
Universit\'e d'Orl\'eans,}
{\mbox Centre National d'\'Etudes Spatiales,}\\
{\mbox 3A Avenue de la Recherche Scientifique, 45071 Orl\'eans, France,}}

\affiliation{\mbox{UFR Sciences et Techniques, 
Universit\'e d’Orl\'eans,}
\mbox{Rue de Chartres, 45100 Orl\'{e}ans, France,}}

\affiliation{\mbox{Observatoire des Sciences de l'Univers en region Centre, UMS 3116,} \\
\mbox{ Universit\'e d'Orl\'eans, Centre National de la Recherche Scientifique, Observatoire de Paris, Universit\'e Paris Sciences \& Lettres,}\\
\mbox{1A rue de la F\'{e}rollerie, 45071 Orl\'{e}ans, France.}}
\date{\today}

\begin{abstract}
The Hubble tension is investigated taking into account the cosmological look-back time. Specifically, considering a single equation, widely used in standard cosmology, it is possible to recover both values of the Hubble constant $H_0$ reported by the SH0ES and Planck collaborations: the former is obtained through cosmological ladder methods (e.g. Cepheids, Supernovae Type IA) and the latter through measurements of the Cosmic Microwave Background. Also, other values obtained in the literature are achieved with the same approach. We conclude that the  Hubble tension can be removed if the look-back time is correctly referred to the redshift where the measurement is performed. 
\end{abstract}



\maketitle
\section{Introduction}
The so-called $H_0$ tension,  related to the reported discrepancy on the measured values of Hubble-Lema\^itre (henceforth Hubble) constant, $H_0$, is claimed to be one of the most critical issues for the $\Lambda$ Cold Dark Matter ($\Lambda$CDM) model, considered the standard picture of the observed Universe. This staggering discrepancy arises when comparing estimates on $H_0$ provided by standard candles with those inferred by the Cosmic Microwave Background  Radiation (CMBR) \cite{Abdalla:2022yfr}. In fact, the results from the SH0ES collaboration, obtained by considering the cosmological ladder and probes like classical Cepheids and Supernovae Type Ia (SNe Ia), in a region with redshift interval  $0.02333<z<0.15$, tell us that $H_0=73.04 \pm 1.04$ km/(s Mpc) at $68 \%$ confidence level (CL) \cite{Riess_2022}, while the Planck collaboration, reporting observations of the CMBR, shows, instead, that $H_0=67.4 \pm 0.5$ km/(s Mpc) at $68 \%$ CL \cite{Planck2020}.
Despite the accuracy of both approaches, a $5.0\sigma$ tension emerges and it seems to be irreconcilable under the standard of the $\Lambda$CDM model.

 The problem has to be addressed both from experimental and theoretical points of view. Considering the increasing precision achieved by recent observations, an experimental reason for the tension, related to some systematic effects, is becoming less and less probable \cite{Abdalla:2022yfr, Salucci2021}. Thus this tension (and other ones revealed in cosmological parameters \cite{Abdalla:2022yfr}) is seemingly hinting at some undefined new physics, going beyond the $\Lambda$CDM model. The last conclusion can be inferred also from other critical issues of the standard cosmological model, like the nature of dark matter and dark energy, which should account for approximately 25\%  and 70\%, respectively,  of the energy-density budget of the observed Universe \cite{riess1998, riess_2007, perlmutter-etal-1999, bahcall-etal-1999, spergel-etal-2003, schimd-etal-2006, mcdonald-etal-2006, bamba-etal-2012, joyce-etal-2015, Salucci2021}. However, the nature of the dark side of the Universe is still unknown at a fundamental level, even if very precise tests are performed by several running experiments \cite{pigozzo-etal-2011,lopezcorredoira-2017, Xenon2017, Workman-etal-2022}.

Different approaches are considered in the literature to fix the above problems. A possibility is taking into account extensions and modifications of General Relativity (GR) where the $\Lambda$CDM model evolves along the cosmic history. For example, extensions like $f(R)$ gravity \cite{Capozziello:2002rd, Nojiri:2010wj, capozziello_2011, Nojiri:2017ncd, DeFelice:2010aj}), have been considered  in cosmological applications \cite{Hu:2007nk, Capozziello:2005ku, Oikonomou:2022irx,Rocco,Bajardi} to address different issues related to the $\Lambda$CDM model, like the late-time dark energy \cite{bamba-etal-2012}, and the inflationary behaviour of the early Universe \cite{Starobinsky1979}. 

Many possible extensions and modifications of GR have been proposed (see e.g. \cite{Cai}), but, no final theory, capable of addressing the whole cosmic history and the large-scale structure has been formulated up to date. As in the case of cosmological inflation, the paradigm seems consistent but a final model, comprehensive of the whole phenomenology, if any, is still missing. In any case, modified gravity can be adopted to investigate the $H_0$ tension, as recently proposed in \cite{Nojiri:2022ski}.

Modified gravity is not the only approach to fix tensions in cosmology. {\it New physics}  can be pursued also by taking into account other paradigms like extensions of Electromagnetism and the Standard Model of Particles \cite{spallicci-etal-2021, Spallicci_2022, Sarracino_2022}, or revising the concept of cosmological measurements according to the principles of Quantum Mechanics  \cite{capozziello-benetti-spallicci-2020, spallicci-benetti-capozziello-2022}, as well as many different other approaches reported in the literature  \cite{Bernal_2016, Mortsell_2018, Vagnozzi_2018, Yang_2018, Poulin_2019, Kreisch_2020, Agrawal_2019, Di_Valentino_2019, Pan_2019, Vagnozzi_2020, Visinelli_2019, Knox_2020, Di_Valentino_2020, Di_Valentino_2020b, Di_Valentino_2021f, Smith_2021, Vagnozzi_2021, Nunes_2021, Cyr_Racine_2022, Anchordoqui_2021, Poulin_2021, Alestas_2022, Smith_2022, Reeves_2023, Poulin_2023, Ester1, Ester2, Ratra_1, Ratra_2}. Furthermore, requiring a phenomenological evolution of $H_0$ with the redshift could be a straightforward way out for the problem \cite{Krishnan_2021, Colgain_2021, Krishnan_2022, Colgain_2022, Colgain_2022b, Colgain_2022c, dainotti_2021a, Dainotti_2022a, Dainotti_2023, Schiavone_2022, Schiavone_2022b, Malekjani_2023, Vahe, Hu_2023} even without investigating some fundamental counterpart.

A more conservative approach is assuming, in any case,  the validity of the $\Lambda$CDM model and the correctness of the reported measurements. In other words, we can investigate the possibility that both $\Lambda$CDM and observations can be somehow reconciled by searching for suitable observables. In other words, the issue of $H_0$ and other tensions could be nothing else but a problem of misinterpretation of measurements at high and low redshift which have to be related in the right way. 

In this paper, we wish to show that the $H_0$ tension can be addressed by the correct interpretation of the {\it look-back time} at various redshifts. It is defined as
\begin{equation} \label{look-back_time}
    T_{lt}(z)=\frac{1}{H_0} \int_0^z \frac{dz'}{(1+z')E(z')}~,
\end{equation}
which expresses the time a photon takes to reach us from a given redshift $z$ in the expanding Universe. It is directly related to the {\it light-travel distance} 
\begin{equation}
d_{lt}(z)=cT_{lt}(z)
\end{equation}
where $c$ is the speed of light.
The key ingredient is the function $E(z)$,  defined as,
\begin{equation} \label{E(z)}
    E(z)=\frac{H(z)}{H_0}=\sqrt{\Omega_r(1+z)^4+\Omega_M(1+z)^3+\Omega_k(1+z)^2+\Omega_{\Lambda}}~,
\end{equation}
the entries of which are the different contributions to the density of the Universe. $\Omega_r$ is the radiation density, $\Omega_M$ the matter density (baryonic and dark), $\Omega_k$ is the density associated with the spatial curvature, and  $\Omega_{\Lambda}$ is the density associated to the cosmological constant. Clearly, $\Omega_{\Lambda}$ can be given by the standard cosmological constant or by some evolving contribution of the Universe vacuum energy. It is worth stressing that Eq. \eqref{E(z)} represents a class of models including  $\Lambda$CDM and its extensions. The role of {\it Precision Cosmology} is to fix consistently the values of $\Omega_i$ parameters as correctly as possible.

Previous works \cite{Jimenez_2019, Bernal_2021, Boylan_Kolchin_2021, Krishnan_2021b, Vagnozzi_2022, Cimatti_2023} have already used the look-back time and the age of the Universe for investigations on the Hubble tension. 

The layout of the paper is the following. Sec. 2  is a general discussion devoted on how to infer $H_0$ from Eq. (\ref{look-back_time}) starting from the age of the Universe, $T_0$, via the same equation valid at any redshift $z$. In Sec. 3, we show how, inserting the values of $T_0, \Omega_r, \Omega_M, \Omega_k,$ and $\Omega_{\Lambda}$,  provided by the Planck collaboration, it is possible to derive values of $H_0$ consistent with both  SH0ES  and  Planck. Also, $H_0$ measurements, reported by other collaborations, are consistent with the presented approach. Conclusions are drawn in Sec. 4.

\section{Look-back time and the Hubble Tension}
The Hubble constant can be inferred by the look-back time at a given redshift by inverting Eq.(\ref{look-back_time}) 
\begin{equation} \label{H0_look-back}
    H_0=\frac{1}{T_{lt}(z)} \int_0^z \frac{dz'}{(1+z')E(z')}~.
\end{equation}
On the other hand, $T_{lt}$ can be linked to the age of the Universe considering the difference between this quantity, evaluated today, and that computed at a given redshift $T(z)$. It is  
\begin{equation} \label{Times}
    T_{lt}=T_0-T(z)=T_0-\frac{T_0}{(1+z)}=T_0-a(t)T_0 ~,
\end{equation}
where we used the definition of the scale factor of the Universe $a(t)$ as
\begin{equation}
    \frac{a_0}{a(t)}=1+z~.
\end{equation}
Here $a_0=1$ is the scale factor corresponding to our epoch\footnote{The parametrization $T(z)=a(t)T_0$ is assumed as a label for the age of the Universe at a given redshift. Thus $a(t)$ is a projection factor and we do not need an integration over the cosmic time.}. In doing so, we can infer the value of the Hubble parameter $H_0$ at any redshift starting from the age of the Universe and the function $E(z)$ from Eq.(\ref{H0_look-back}).  It is
\begin{equation} \label{H0_Riess}
    H_0^{(z)}=\frac{(1+z)}{T_0 z} \int_0^z \frac{dz'}{(1+z')E(z')}~.
\end{equation}
We notice that, for $z \to +\infty$, it is
$
    {\displaystyle \lim_{z \to +\infty} \frac{z+1}{z} = 1,}
$
and thus 
\begin{equation} \label{H0_Planck}
     H_0^{(\infty)} =\frac{1}{T_0} \int_0^\infty \frac{dz'}{(1+z')E(z')}~.
\end{equation}
Eqs. (\ref{H0_Riess}) and  (\ref{H0_Planck})  tell us that it is possible to find different values for $H_0$ at different values of the redshift, and that the $H_0$ tension could be addressed by reconciling the value of $H_0$, derived at a given  $z$,  with the one derived at $z \to +\infty$. In this perspective, we may conclude that the $H_0$ tension is not due to neither systematic errors, nor to some new physics beyond the $\Lambda$CDM model, but simply to the fact that, in measuring the same quantity at different redshift, different results arise thanks to the look-back time evaluated at different epochs.

\section{Recovering $H_0$ from  Planck and SH0ES results}
Let us now test the above considerations by adopting observational results provided by the Planck and SH0ES collaborations. The claim is that  Eq. (\ref{H0_Riess}) gives  $H_0$ reported in \cite{Riess_2022}
and Eq. (\ref{H0_Planck}) gives the one reported in \cite{Planck2020}.
According to the latter reference, the values that fix the $\Lambda$CDM model are
\begin{equation}
    T_0=13.797 \text{Gyr}, \quad \Omega_r=9.252 \times 10^{-5},\quad  \Omega_M=0.3153\,, \quad \Omega_{\Lambda}=0.6847\,.
\end{equation}
Taking into account these values, and fixing $z=0.15$ (the upper limit of the redshift interval of the observations performed by the SH0ES collaboration), we find, from Eq. (\ref{H0_Riess})
\begin{equation} \label{Riess_result}
    H_0^S=73.29 \quad \text{km/(s Mpc)} ~,
\end{equation}
which is remarkably consistent with the value inferred in Ref. \cite{Riess_2022}.  It is worth noticing that this result is achieved by using values provided by the Planck collaboration, further proving the concordance between the two results. Furthermore,  from Eq. \eqref{H0_Planck}, we find
\begin{equation} \label{Planck_result}
    H_0^P=67.40 \quad \text{km/(s Mpc)} ~,
\end{equation}
which is consistent with the $H_0$ value reported by the Planck collaboration. In both cases, $\Lambda$CDM has been adopted but the look-back time has been considered at redshifts consistent with observational data.

\begin{figure}
\includegraphics[width=0.8\hsize,height=0.6\textwidth]{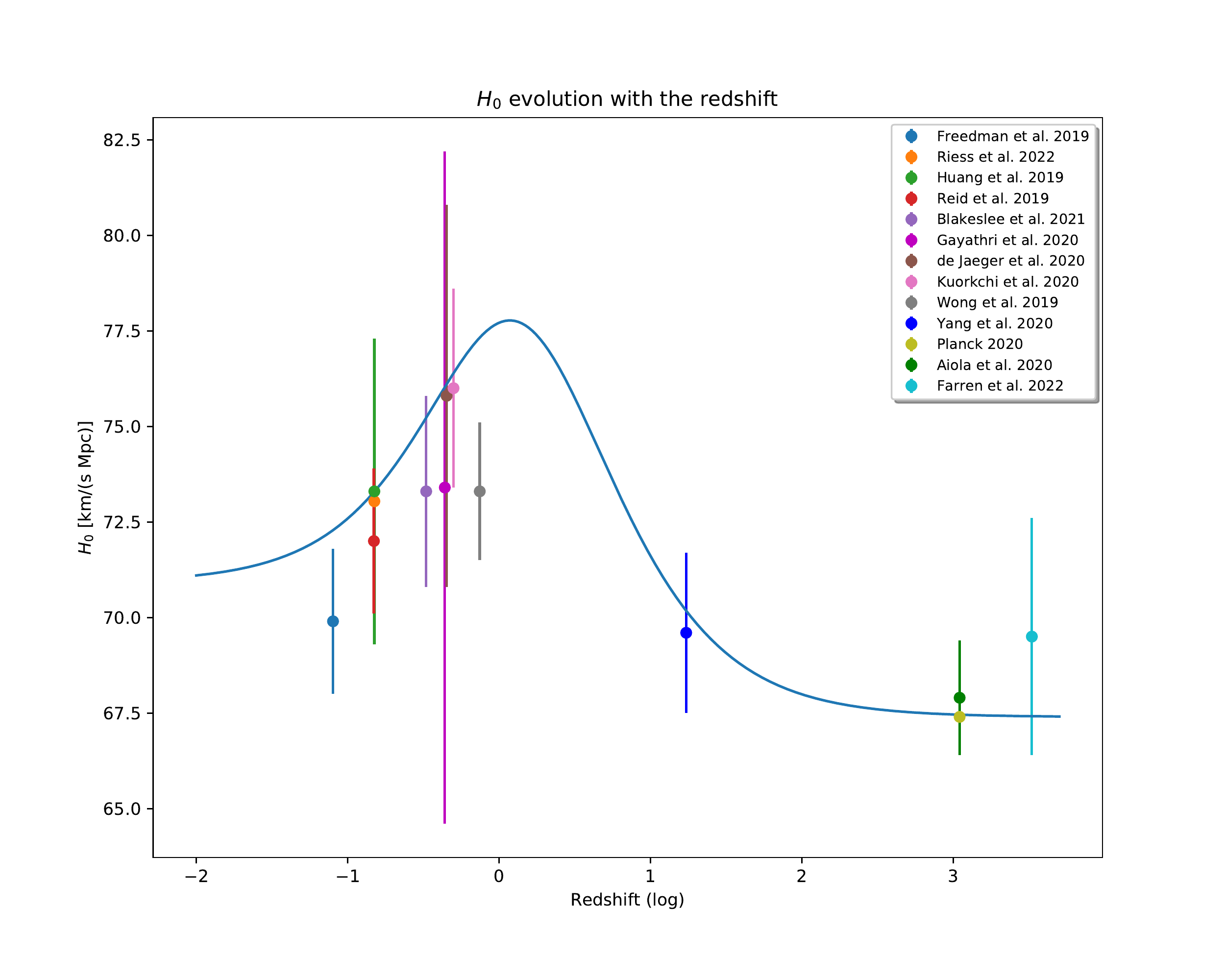}
\caption{The value of $H_0$ derived from Eq.(\ref{H0_Riess}) vs the redshift, confronted with observational data. The values for the redshifts representing each measurement have been chosen either at the upper limit of the redshift interval sample, or at the redshift corresponding to the given phenomenon considered for the estimate (Cosmic Dawn, Recombination, Equivalence).  The $x$-axis is in logarithmic scale.}
\label{Figure_1}
\end{figure}

The approach can be generalised to any redshift, finding how $H_0$ changes as a function of the redshift. In Fig.\ref{Figure_1},  we plot Eq.(\ref{H0_Riess}) confronting the values of $H_0$ reported by various collaborations. We choose the upper limit of the $z$ interval reported in the related literature, or the redshift adopted to estimate $H_0$ corresponding to a given physical phenomenon. 
Besides the $H_0$ results,  provided by \cite{Riess_2022} and \cite{Planck2020}, we report on the plot the following measurements, all at $68 \%$ CL:
\begin{itemize}

\item $H_0=69.9 \pm 1.9\, \text{km/(s Mpc)}$, obtained by using the tip of the red giants branch as an anchor for SNe Ia, at $z=0.08$ \cite{Freedman_2019};

\item $H_0=75.8 \pm 5.0\, \text{km/(s Mpc)}$, derived from SNe Type II, at $z=0.45$ \cite{de_Jaeger_2020}; 

\item $H_0=73.3 \pm 4.0\, \text{km/(s Mpc)}$, derived from the Mira Variables and SNe Ia,  at $z \sim 0.15$ \cite{Huang_2020}; 

\item $H_0=76.0 \pm 2.6\, \text{km/(s Mpc)}$, derived from the Tully-Fisher relation, at $z=0.5$ \cite{Kourkchi_2020}; 

\item $H_0=73.3 \pm 2.5\, \text{km/(s Mpc)}$, derived from the surface brightness fluctuations, at $z=0.33$ \cite{Blakeslee_2021}; 

\item $H_0=69.5 \pm 3.3\, \text{km/(s Mpc)}$, inferred from the Large Scale Structure $t_{eq}$ standard ruler, and thus confronted to our computations at the redshift of equivalence $z_{eq} \sim 3300$ \cite{Farren_2022}; 

\item $H_0=72.0 \pm 1.9\, \text{km/(s Mpc)}$, inferred from the masers+ SNe Ia and compared at $z \sim 0.15$ \cite{Reid_2019}; 

\item $H_0=73.3 \pm 1.8\, \text{km/(s Mpc)}$, derived from gravitational lensed quasars, confronted with our plot at $z=0.745$; 

\item $H_0=67.9 \pm 1.5\, \text{km/(s Mpc)}$, which is a measurement obtained by the CMBR independently from the Planck collaboration, and thus corresponding, in our plot, at the reionization epoch $z \sim 1100$ \cite{Aiola_2020}; 

\item $H_0=69.6 \pm 2.1\, \text{km/(s Mpc)}$, linked to the 21 cm absorption line and corresponding at the beginning of the so-called Cosmic Dawn, the epoch when the first stars formed ($z \sim 17.2$), in combination with CMBR data and considering a Chaplygin gas model for the dark sector \cite{Yang_2019}; 

\item Finally, $H_0=73.4\pm 8.8\, \text{km/(s Mpc)}$, deduced by gravitational waves, at $z=0.438$ \cite{Gayathri_2020}.

\end{itemize}

From  Fig. \ref{Figure_1}, it is clear that most of the $H_0$ measurements are consistent, within 1$\sigma$,  with respect to the curve obtained by  Eq. (\ref{H0_Riess}) for any $z$. The two outliers, corresponding to the measurements reported in  \cite{Freedman_2019, Wong_2019}, are still consistent within 2$\sigma$. It is interesting to see that the trend of $H_0$ values matches Eq. (\ref{H0_Riess}) also if different methods and cosmological probes have been adopted. This means that any $H_0$ measurement can be somehow related to the look-back time.
We comment that, according to  Fig. \ref{Figure_1}, $H_0$ measurements in the range $z \sim 2\div 3$ ($0.30 \div 0.48$ in log) are expected to show even higher values for this quantity, and thus a stronger tension with respect to the Planck results. Cosmological observations and computations in this intermediate redshift range exist but they strictly depend on calibrations at lower redshift \cite{scolnic-2018, Scolnic_2022, Dainotti_2022e, Dainotti_2022f, Bargiacchi_2022}. Thus, new observations are expected by novel probes, possibly providing a standalone estimate of cosmological parameters in this redshift region, like quasars, novel gravitational wave standard sirens, galaxy clusters, Lyman-$\alpha$ lines, or gamma-ray bursts \cite{Dainotti_2022c,Califano1, Califano2,Bargiacchi_3}.

\section{Discussion and Conclusions}

The problem of the $H_0$ tension can be addressed in the framework of the $\Lambda$CDM model in terms of look-back time considered at the redshift where measurements are performed. In this derivation, the values of the age of the Universe and the cosmic densities, provided by the Planck collaboration, are adopted. The look-back time curve, matching the observational values of $H_0$, is shown in Fig. \ref{Figure_1}. According to this approach, we are able to recover most of the reported $H_0$ measurements and, more remarkably, the two most representatives for the $H_0$ tension, namely the ones provided by the Planck \cite{Planck2020} and SH0ES \cite{Riess_2022} collaborations. The result is interesting because the concordance of $H_0$ measurements is achieved independently from the type of performed measurement. Indeed, we have considered estimates obtained by different standard candles (SNe Ia, TRGB, Cepheids, etc), phenomena concerning early stages of the Universe (CMBR,  Cosmic Dawn,  Equivalence Epoch), and also novel probes (gravitational waves). According to this picture, we may conclude that no systematic errors nor new physics are necessary to explain the tension, which, instead, can be accounted for with a single formula in the framework of the $\Lambda$CDM model. In other words, by adopting the look-back time, we can consistently account for late-type and early-type measurements, confirming the underlying consistency of the $\Lambda$CDM model. We also point out that this evolution of $H_0$ constant in the framework of the $\Lambda$CDM model does not change its role as a formal integration constant derived from the Friedman equations, but its operative value depends on the redshift where it is inferred. This conclusion is important for the definition of the cosmological distances, which are all depending on $H_0$.

It is worth stressing that the value of the age of the Universe, $T_0$, has been derived by the Planck collaboration starting from six independent parameters (the minimum number of parameters necessary to describe the $\Lambda$CDM model), among which  $H_0$ is not included. In fact, also $H_0$ is inferred from the same minimum number of independent parameters. This means that any circularity problem is avoided.
Furthermore, we can note that the plot in Fig. \ref{Figure_1} is in agreement with the general trend of $H_0$  late-type measurements. See, for example, Ref. \cite{Di_Valentino_2021} where the estimate $H_0=72.94 \pm 0.75 \text{km/(s Mpc)}$ at $68 \%$ CL is reported.
Furthermore, our results are consistent with constraints on $H_0$ obtained from the age of the Universe: for instance, in \cite{Vagnozzi_2021}, an upper limit at $H_0=73.2 \; \text{km/(s Mpc)}$ with $95 \%$ CL has been inferred by considering the ages of old astrophysical objects up to $z \sim 8$. This value is consistent with Fig. \ref{Figure_1}, because, according to our approach, we find $H_0=72.3 \; \text{km/(s Mpc)}$ for that particular value of the redshift.

Finally, computations reported here are consistent with the interpretation proposed in \cite{capozziello-benetti-spallicci-2020, spallicci-benetti-capozziello-2022}, according to which a fundamental uncertainty on $H_0$ exists {\it a priori}  in the framework of the $\Lambda$CDM scenario. Furthermore,  by the light-travel distance, a  cosmological  Compton length can be always defined.  This distance is strictly connected to the look-back time \cite{capozziello-benetti-spallicci-2020} without contradiction with present results.

As a final remark, it is useful noticing that we do not exclude the idea of new physics beyond the $\Lambda$CDM model, because other issues, like dark energy and dark matter, at late time, or inflation, at early time, could require to go beyond GR and standard cosmology. However, from the above discussion, it is possible to state that new physics is not necessary for the particular issue of the $H_0$ tension, which can be completely accounted for in the framework of the $\Lambda$CDM model.

\section*{Aknowledgements}
This paper is based upon work from the COST Action CA21136 Addressing observational tensions in cosmology with systematics and fundamental physics (CosmoVerse) supported by COST (European Cooperation in Science and Technology). SC and GS are grateful to the LPC2E laboratory, {\it Universit\'e d'Orl\'eans,} for the hospitality while writing this work. They also acknowledge the support of Istituto Nazionale di Fisica Nucleare, Sez. di Napoli, Iniziative Specifiche MOONLIGHT-2 and QGSKY.

\bibliography{mybibliographyHubble}

\begin{thebibliography}{105}
\expandafter\ifx\csname natexlab\endcsname\relax\def\natexlab#1{#1}\fi
\expandafter\ifx\csname bibnamefont\endcsname\relax
  \def\bibnamefont#1{#1}\fi
\expandafter\ifx\csname bibfnamefont\endcsname\relax
  \def\bibfnamefont#1{#1}\fi
\expandafter\ifx\csname citenamefont\endcsname\relax
  \def\citenamefont#1{#1}\fi
\expandafter\ifx\csname url\endcsname\relax
  \def\url#1{\texttt{#1}}\fi
\expandafter\ifx\csname urlprefix\endcsname\relax\def\urlprefix{URL }\fi
\providecommand{\bibinfo}[2]{#2}
\providecommand{\eprint}[2][]{\url{#2}}

\bibitem[{\citenamefont{Abdalla et~al.}(2022)\citenamefont{Abdalla,
  Abell{\'{a}}n, Aboubrahim, Agnello, Özgür Akarsu, Akrami, Alestas, Aloni,
  Amendola, Anchordoqui et~al.}}]{Abdalla:2022yfr}
\bibinfo{author}{\bibfnamefont{E.}~\bibnamefont{Abdalla}},
  \bibinfo{author}{\bibfnamefont{G.~F.} \bibnamefont{Abell{\'{a}}n}},
  \bibinfo{author}{\bibfnamefont{A.}~\bibnamefont{Aboubrahim}},
  \bibinfo{author}{\bibfnamefont{A.}~\bibnamefont{Agnello}},
  \bibinfo{author}{\bibnamefont{Özgür Akarsu}},
  \bibinfo{author}{\bibfnamefont{Y.}~\bibnamefont{Akrami}},
  \bibinfo{author}{\bibfnamefont{G.}~\bibnamefont{Alestas}},
  \bibinfo{author}{\bibfnamefont{D.}~\bibnamefont{Aloni}},
  \bibinfo{author}{\bibfnamefont{L.}~\bibnamefont{Amendola}},
  \bibinfo{author}{\bibfnamefont{L.~A.} \bibnamefont{Anchordoqui}},
  \bibnamefont{et~al.}, \bibinfo{journal}{Journal of High Energy Astrophysics}
  \textbf{\bibinfo{volume}{34}}, \bibinfo{pages}{49} (\bibinfo{year}{2022}),
  \urlprefix\url{https://doi.org/10.1016/j.jheap.2022.04.002}.

\bibitem[{\citenamefont{Riess et~al.}(2022)\citenamefont{Riess, Yuan, Macri,
  Scolnic, Brout, Casertano, Jones, Murakami, Anand, Breuval
  et~al.}}]{Riess_2022}
\bibinfo{author}{\bibfnamefont{A.~G.} \bibnamefont{Riess}},
  \bibinfo{author}{\bibfnamefont{W.}~\bibnamefont{Yuan}},
  \bibinfo{author}{\bibfnamefont{L.~M.} \bibnamefont{Macri}},
  \bibinfo{author}{\bibfnamefont{D.}~\bibnamefont{Scolnic}},
  \bibinfo{author}{\bibfnamefont{D.}~\bibnamefont{Brout}},
  \bibinfo{author}{\bibfnamefont{S.}~\bibnamefont{Casertano}},
  \bibinfo{author}{\bibfnamefont{D.~O.} \bibnamefont{Jones}},
  \bibinfo{author}{\bibfnamefont{Y.}~\bibnamefont{Murakami}},
  \bibinfo{author}{\bibfnamefont{G.~S.} \bibnamefont{Anand}},
  \bibinfo{author}{\bibfnamefont{L.}~\bibnamefont{Breuval}},
  \bibnamefont{et~al.}, \bibinfo{journal}{The Astrophysical Journal Letters}
  \textbf{\bibinfo{volume}{934}}, \bibinfo{pages}{L7} (\bibinfo{year}{2022}),
  \urlprefix\url{https://doi.org/10.3847/2041-8213/ac5c5b}.

\bibitem[{\citenamefont{Aghanim et~al.}(2020)\citenamefont{Aghanim, Akrami,
  Ashdown, Aumont, Baccigalupi, Ballardini, Banday, Barreiro, Bartolo, Basak
  et~al.}}]{Planck2020}
\bibinfo{author}{\bibfnamefont{N.}~\bibnamefont{Aghanim}},
  \bibinfo{author}{\bibfnamefont{Y.}~\bibnamefont{Akrami}},
  \bibinfo{author}{\bibfnamefont{M.}~\bibnamefont{Ashdown}},
  \bibinfo{author}{\bibfnamefont{J.}~\bibnamefont{Aumont}},
  \bibinfo{author}{\bibfnamefont{C.}~\bibnamefont{Baccigalupi}},
  \bibinfo{author}{\bibfnamefont{M.}~\bibnamefont{Ballardini}},
  \bibinfo{author}{\bibfnamefont{A.~J.} \bibnamefont{Banday}},
  \bibinfo{author}{\bibfnamefont{R.~B.} \bibnamefont{Barreiro}},
  \bibinfo{author}{\bibfnamefont{N.}~\bibnamefont{Bartolo}},
  \bibinfo{author}{\bibfnamefont{S.}~\bibnamefont{Basak}},
  \bibnamefont{et~al.}, \bibinfo{journal}{Astronomy \& Astrophysics}
  \textbf{\bibinfo{volume}{641}}, \bibinfo{pages}{A6} (\bibinfo{year}{2020}),
  \urlprefix\url{http://dx.doi.org/10.1051/0004-6361/201833910}.

\bibitem[{\citenamefont{Salucci et~al.}(2021)\citenamefont{Salucci, Esposito,
  Lambiase, Battista, Benetti, Bini, Boco, Sharma, Bozza, Buoninfante
  et~al.}}]{Salucci2021}
\bibinfo{author}{\bibfnamefont{P.}~\bibnamefont{Salucci}},
  \bibinfo{author}{\bibfnamefont{G.}~\bibnamefont{Esposito}},
  \bibinfo{author}{\bibfnamefont{G.}~\bibnamefont{Lambiase}},
  \bibinfo{author}{\bibfnamefont{E.}~\bibnamefont{Battista}},
  \bibinfo{author}{\bibfnamefont{M.}~\bibnamefont{Benetti}},
  \bibinfo{author}{\bibfnamefont{D.}~\bibnamefont{Bini}},
  \bibinfo{author}{\bibfnamefont{L.}~\bibnamefont{Boco}},
  \bibinfo{author}{\bibfnamefont{G.}~\bibnamefont{Sharma}},
  \bibinfo{author}{\bibfnamefont{V.}~\bibnamefont{Bozza}},
  \bibinfo{author}{\bibfnamefont{L.}~\bibnamefont{Buoninfante}},
  \bibnamefont{et~al.}, \bibinfo{journal}{Frontiers in Physics}
  \textbf{\bibinfo{volume}{8}} (\bibinfo{year}{2021}),
  \urlprefix\url{http://dx.doi.org/10.3389/fphy.2020.603190}.

\bibitem[{\citenamefont{Riess et~al.}(1998)\citenamefont{Riess, Filippenko,
  Challis, Clocchiatti, Diercks, Garnavich, Gilliland, Hogan, Jha, Kirshner
  et~al.}}]{riess1998}
\bibinfo{author}{\bibfnamefont{A.~G.} \bibnamefont{Riess}},
  \bibinfo{author}{\bibfnamefont{A.~V.} \bibnamefont{Filippenko}},
  \bibinfo{author}{\bibfnamefont{P.}~\bibnamefont{Challis}},
  \bibinfo{author}{\bibfnamefont{A.}~\bibnamefont{Clocchiatti}},
  \bibinfo{author}{\bibfnamefont{A.}~\bibnamefont{Diercks}},
  \bibinfo{author}{\bibfnamefont{P.~M.} \bibnamefont{Garnavich}},
  \bibinfo{author}{\bibfnamefont{R.~L.} \bibnamefont{Gilliland}},
  \bibinfo{author}{\bibfnamefont{C.~J.} \bibnamefont{Hogan}},
  \bibinfo{author}{\bibfnamefont{S.}~\bibnamefont{Jha}},
  \bibinfo{author}{\bibfnamefont{R.~P.} \bibnamefont{Kirshner}},
  \bibnamefont{et~al.}, \bibinfo{journal}{The Astronomical Journal}
  \textbf{\bibinfo{volume}{116}}, \bibinfo{pages}{1009} (\bibinfo{year}{1998}),
  \urlprefix\url{http://dx.doi.org/10.1086/300499}.

\bibitem[{\citenamefont{Riess et~al.}(2007)\citenamefont{Riess, Strolger,
  Casertano, Ferguson, Mobasher, Gold, Challis, Filippenko, Jha, Li
  et~al.}}]{riess_2007}
\bibinfo{author}{\bibfnamefont{A.~G.} \bibnamefont{Riess}},
  \bibinfo{author}{\bibfnamefont{L.-G.} \bibnamefont{Strolger}},
  \bibinfo{author}{\bibfnamefont{S.}~\bibnamefont{Casertano}},
  \bibinfo{author}{\bibfnamefont{H.~C.} \bibnamefont{Ferguson}},
  \bibinfo{author}{\bibfnamefont{B.}~\bibnamefont{Mobasher}},
  \bibinfo{author}{\bibfnamefont{B.}~\bibnamefont{Gold}},
  \bibinfo{author}{\bibfnamefont{P.~J.} \bibnamefont{Challis}},
  \bibinfo{author}{\bibfnamefont{A.~V.} \bibnamefont{Filippenko}},
  \bibinfo{author}{\bibfnamefont{S.}~\bibnamefont{Jha}},
  \bibinfo{author}{\bibfnamefont{W.}~\bibnamefont{Li}}, \bibnamefont{et~al.},
  \bibinfo{journal}{The Astrophysical Journal} \textbf{\bibinfo{volume}{659}},
  \bibinfo{pages}{98} (\bibinfo{year}{2007}),
  \urlprefix\url{https://doi.org/10.1086/510378}.

\bibitem[{\citenamefont{Perlmutter et~al.}(1999)\citenamefont{Perlmutter,
  Aldering, Goldhaber, Knop, Nugent, Castro, Deustua, Fabbro, Goobar, Groom
  et~al.}}]{perlmutter-etal-1999}
\bibinfo{author}{\bibfnamefont{S.}~\bibnamefont{Perlmutter}},
  \bibinfo{author}{\bibfnamefont{G.}~\bibnamefont{Aldering}},
  \bibinfo{author}{\bibfnamefont{G.}~\bibnamefont{Goldhaber}},
  \bibinfo{author}{\bibfnamefont{R.~A.} \bibnamefont{Knop}},
  \bibinfo{author}{\bibfnamefont{P.}~\bibnamefont{Nugent}},
  \bibinfo{author}{\bibfnamefont{P.~G.} \bibnamefont{Castro}},
  \bibinfo{author}{\bibfnamefont{S.}~\bibnamefont{Deustua}},
  \bibinfo{author}{\bibfnamefont{S.}~\bibnamefont{Fabbro}},
  \bibinfo{author}{\bibfnamefont{A.}~\bibnamefont{Goobar}},
  \bibinfo{author}{\bibfnamefont{D.~E.} \bibnamefont{Groom}},
  \bibnamefont{et~al.}, \bibinfo{journal}{The Astrophysical Journal}
  \textbf{\bibinfo{volume}{517}}, \bibinfo{pages}{565} (\bibinfo{year}{1999}),
  \urlprefix\url{https://doi.org/10.1086\%2F307221}.

\bibitem[{\citenamefont{Bahcall et~al.}(1999)\citenamefont{Bahcall, Ostriker,
  Perlmutter, and Steinhardt}}]{bahcall-etal-1999}
\bibinfo{author}{\bibfnamefont{N.~A.} \bibnamefont{Bahcall}},
  \bibinfo{author}{\bibfnamefont{J.~P.} \bibnamefont{Ostriker}},
  \bibinfo{author}{\bibfnamefont{S.}~\bibnamefont{Perlmutter}},
  \bibnamefont{and} \bibinfo{author}{\bibfnamefont{P.~J.}
  \bibnamefont{Steinhardt}}, \bibinfo{journal}{Science}
  \textbf{\bibinfo{volume}{284}}, \bibinfo{pages}{1481} (\bibinfo{year}{1999}),
  \urlprefix\url{https://doi.org/10.1126/science.284.5419.1481}.

\bibitem[{\citenamefont{Spergel et~al.}(2003)\citenamefont{Spergel, Verde,
  Peiris, Komatsu, Nolta, Bennett, Halpern, Hinshaw, Jarosik, Kogut
  et~al.}}]{spergel-etal-2003}
\bibinfo{author}{\bibfnamefont{D.~N.} \bibnamefont{Spergel}},
  \bibinfo{author}{\bibfnamefont{L.}~\bibnamefont{Verde}},
  \bibinfo{author}{\bibfnamefont{H.~V.} \bibnamefont{Peiris}},
  \bibinfo{author}{\bibfnamefont{E.}~\bibnamefont{Komatsu}},
  \bibinfo{author}{\bibfnamefont{M.~R.} \bibnamefont{Nolta}},
  \bibinfo{author}{\bibfnamefont{C.~L.} \bibnamefont{Bennett}},
  \bibinfo{author}{\bibfnamefont{M.}~\bibnamefont{Halpern}},
  \bibinfo{author}{\bibfnamefont{G.}~\bibnamefont{Hinshaw}},
  \bibinfo{author}{\bibfnamefont{N.}~\bibnamefont{Jarosik}},
  \bibinfo{author}{\bibfnamefont{A.}~\bibnamefont{Kogut}},
  \bibnamefont{et~al.}, \bibinfo{journal}{The Astrophysical Journal Supplement
  Series} \textbf{\bibinfo{volume}{148}}, \bibinfo{pages}{175}
  (\bibinfo{year}{2003}), \urlprefix\url{https://doi.org/10.1086/377226}.

\bibitem[{\citenamefont{Schimd et~al.}(2006)\citenamefont{Schimd, Tereno, Uzan,
  Mellier, van Waerbeke, Semboloni, Hoekstra, Fu, and
  Riazuelo}}]{schimd-etal-2006}
\bibinfo{author}{\bibfnamefont{C.}~\bibnamefont{Schimd}},
  \bibinfo{author}{\bibfnamefont{I.}~\bibnamefont{Tereno}},
  \bibinfo{author}{\bibfnamefont{J.-P.} \bibnamefont{Uzan}},
  \bibinfo{author}{\bibfnamefont{Y.}~\bibnamefont{Mellier}},
  \bibinfo{author}{\bibfnamefont{L.}~\bibnamefont{van Waerbeke}},
  \bibinfo{author}{\bibfnamefont{E.}~\bibnamefont{Semboloni}},
  \bibinfo{author}{\bibfnamefont{H.}~\bibnamefont{Hoekstra}},
  \bibinfo{author}{\bibfnamefont{L.}~\bibnamefont{Fu}}, \bibnamefont{and}
  \bibinfo{author}{\bibfnamefont{A.}~\bibnamefont{Riazuelo}},
  \bibinfo{journal}{Astronomy \& Astrophysics} \textbf{\bibinfo{volume}{463}},
  \bibinfo{pages}{405} (\bibinfo{year}{2006}),
  \urlprefix\url{https://doi.org/10.1051/0004-6361:20065154}.

\bibitem[{\citenamefont{McDonald et~al.}(2006)\citenamefont{McDonald, Seljak,
  Burles, Schlegel, Weinberg, Cen, Shih, Schaye, Schneider, Bahcall
  et~al.}}]{mcdonald-etal-2006}
\bibinfo{author}{\bibfnamefont{P.}~\bibnamefont{McDonald}},
  \bibinfo{author}{\bibfnamefont{U.}~\bibnamefont{Seljak}},
  \bibinfo{author}{\bibfnamefont{S.}~\bibnamefont{Burles}},
  \bibinfo{author}{\bibfnamefont{D.~J.} \bibnamefont{Schlegel}},
  \bibinfo{author}{\bibfnamefont{D.~H.} \bibnamefont{Weinberg}},
  \bibinfo{author}{\bibfnamefont{R.}~\bibnamefont{Cen}},
  \bibinfo{author}{\bibfnamefont{D.}~\bibnamefont{Shih}},
  \bibinfo{author}{\bibfnamefont{J.}~\bibnamefont{Schaye}},
  \bibinfo{author}{\bibfnamefont{D.~P.} \bibnamefont{Schneider}},
  \bibinfo{author}{\bibfnamefont{N.~A.} \bibnamefont{Bahcall}},
  \bibnamefont{et~al.}, \bibinfo{journal}{The Astrophysical Journal Supplement
  Series} \textbf{\bibinfo{volume}{163}}, \bibinfo{pages}{80}
  (\bibinfo{year}{2006}), \urlprefix\url{https://doi.org/10.1086/444361}.

\bibitem[{\citenamefont{Bamba et~al.}(2012)\citenamefont{Bamba, Capozziello,
  Nojiri, and Odintsov}}]{bamba-etal-2012}
\bibinfo{author}{\bibfnamefont{K.}~\bibnamefont{Bamba}},
  \bibinfo{author}{\bibfnamefont{S.}~\bibnamefont{Capozziello}},
  \bibinfo{author}{\bibfnamefont{S.}~\bibnamefont{Nojiri}}, \bibnamefont{and}
  \bibinfo{author}{\bibfnamefont{S.~D.} \bibnamefont{Odintsov}},
  \bibinfo{journal}{Astrophysics and Space Science}
  \textbf{\bibinfo{volume}{342}}, \bibinfo{pages}{155} (\bibinfo{year}{2012}),
  \urlprefix\url{https://doi.org/10.1007/s10509-012-1181-8}.

\bibitem[{\citenamefont{Joyce et~al.}(2015)\citenamefont{Joyce, Jain, Khoury,
  and Trodden}}]{joyce-etal-2015}
\bibinfo{author}{\bibfnamefont{A.}~\bibnamefont{Joyce}},
  \bibinfo{author}{\bibfnamefont{B.}~\bibnamefont{Jain}},
  \bibinfo{author}{\bibfnamefont{J.}~\bibnamefont{Khoury}}, \bibnamefont{and}
  \bibinfo{author}{\bibfnamefont{M.}~\bibnamefont{Trodden}},
  \bibinfo{journal}{Physics Reports} \textbf{\bibinfo{volume}{568}},
  \bibinfo{pages}{1} (\bibinfo{year}{2015}),
  \urlprefix\url{https://doi.org/10.1016/j.physrep.2014.12.002}.

\bibitem[{\citenamefont{Pigozzo et~al.}(2011)\citenamefont{Pigozzo, Dantas,
  Carneiro, and Alcaniz}}]{pigozzo-etal-2011}
\bibinfo{author}{\bibfnamefont{C.}~\bibnamefont{Pigozzo}},
  \bibinfo{author}{\bibfnamefont{M.}~\bibnamefont{Dantas}},
  \bibinfo{author}{\bibfnamefont{S.}~\bibnamefont{Carneiro}}, \bibnamefont{and}
  \bibinfo{author}{\bibfnamefont{J.}~\bibnamefont{Alcaniz}},
  \bibinfo{journal}{Journal of Cosmology and Astroparticle Physics}
  \textbf{\bibinfo{volume}{2011}}, \bibinfo{pages}{022} (\bibinfo{year}{2011}),
  \urlprefix\url{https://doi.org/10.1088/1475-7516/2011/08/022}.

\bibitem[{\citenamefont{L{\'{o}}pez-Corredoira}(2017)}]{lopezcorredoira-2017}
\bibinfo{author}{\bibfnamefont{M.}~\bibnamefont{L{\'{o}}pez-Corredoira}},
  \bibinfo{journal}{Foundations of Physics} \textbf{\bibinfo{volume}{47}},
  \bibinfo{pages}{711} (\bibinfo{year}{2017}),
  \urlprefix\url{https://doi.org/10.1007/s10701-017-0073-8}.

\bibitem[{\citenamefont{Aprile et~al.}(2017)\citenamefont{Aprile, Aalbers,
  Agostini, Alfonsi, Amaro, Anthony, Antunes, Arneodo, Balata, Barrow
  et~al.}}]{Xenon2017}
\bibinfo{author}{\bibfnamefont{E.}~\bibnamefont{Aprile}},
  \bibinfo{author}{\bibfnamefont{J.}~\bibnamefont{Aalbers}},
  \bibinfo{author}{\bibfnamefont{F.}~\bibnamefont{Agostini}},
  \bibinfo{author}{\bibfnamefont{M.}~\bibnamefont{Alfonsi}},
  \bibinfo{author}{\bibfnamefont{F.~D.} \bibnamefont{Amaro}},
  \bibinfo{author}{\bibfnamefont{M.}~\bibnamefont{Anthony}},
  \bibinfo{author}{\bibfnamefont{B.}~\bibnamefont{Antunes}},
  \bibinfo{author}{\bibfnamefont{F.}~\bibnamefont{Arneodo}},
  \bibinfo{author}{\bibfnamefont{M.}~\bibnamefont{Balata}},
  \bibinfo{author}{\bibfnamefont{P.}~\bibnamefont{Barrow}},
  \bibnamefont{et~al.}, \bibinfo{journal}{The European Physical Journal C}
  \textbf{\bibinfo{volume}{77}}, \bibinfo{pages}{881} (\bibinfo{year}{2017}),
  \urlprefix\url{http://dx.doi.org/10.1140/epjc/s10052-017-5326-3}.

\bibitem[{\citenamefont{Workman and {the Particle Data
  Group}}(2022)}]{Workman-etal-2022}
\bibinfo{author}{\bibfnamefont{R.~L.} \bibnamefont{Workman}} \bibnamefont{and}
  \bibinfo{author}{\bibnamefont{{the Particle Data Group}}},
  \bibinfo{journal}{Progress of Theoretical and Experimental Physics} p.
  \bibinfo{pages}{083C01} (\bibinfo{year}{2022}),
  \urlprefix\url{https://doi.org/10.1093/ptep/ptac097}.

\bibitem[{\citenamefont{Capozziello}(2002)}]{Capozziello:2002rd}
\bibinfo{author}{\bibfnamefont{S.}~\bibnamefont{Capozziello}},
  \bibinfo{journal}{International Journal of Modern Physics D}
  \textbf{\bibinfo{volume}{11}}, \bibinfo{pages}{483} (\bibinfo{year}{2002}),
  \urlprefix\url{https://doi.org/10.1142/s0218271802002025}.

\bibitem[{\citenamefont{Nojiri and Odintsov}(2011)}]{Nojiri:2010wj}
\bibinfo{author}{\bibfnamefont{S.}~\bibnamefont{Nojiri}} \bibnamefont{and}
  \bibinfo{author}{\bibfnamefont{S.~D.} \bibnamefont{Odintsov}},
  \bibinfo{journal}{Physics Reports} \textbf{\bibinfo{volume}{505}},
  \bibinfo{pages}{59} (\bibinfo{year}{2011}),
  \urlprefix\url{https://doi.org/10.1016/j.physrep.2011.04.001}.

\bibitem[{\citenamefont{Capozziello et~al.}(2011)\citenamefont{Capozziello,
  Laurentis, Odintsov, and Stabile}}]{capozziello_2011}
\bibinfo{author}{\bibfnamefont{S.}~\bibnamefont{Capozziello}},
  \bibinfo{author}{\bibfnamefont{M.~D.} \bibnamefont{Laurentis}},
  \bibinfo{author}{\bibfnamefont{S.}~\bibnamefont{Odintsov}}, \bibnamefont{and}
  \bibinfo{author}{\bibfnamefont{A.}~\bibnamefont{Stabile}},
  \bibinfo{journal}{Physical Review D} \textbf{\bibinfo{volume}{83}}
  (\bibinfo{year}{2011}),
  \urlprefix\url{https://doi.org/10.1103/physrevd.83.064004}.

\bibitem[{\citenamefont{Nojiri et~al.}(2017)\citenamefont{Nojiri, Odintsov, and
  Oikonomou}}]{Nojiri:2017ncd}
\bibinfo{author}{\bibfnamefont{S.}~\bibnamefont{Nojiri}},
  \bibinfo{author}{\bibfnamefont{S.}~\bibnamefont{Odintsov}}, \bibnamefont{and}
  \bibinfo{author}{\bibfnamefont{V.}~\bibnamefont{Oikonomou}},
  \bibinfo{journal}{Physics Reports} \textbf{\bibinfo{volume}{692}},
  \bibinfo{pages}{1} (\bibinfo{year}{2017}),
  \urlprefix\url{https://doi.org/10.1016/j.physrep.2017.06.001}.

\bibitem[{\citenamefont{Felice and Tsujikawa}(2010)}]{DeFelice:2010aj}
\bibinfo{author}{\bibfnamefont{A.~D.} \bibnamefont{Felice}} \bibnamefont{and}
  \bibinfo{author}{\bibfnamefont{S.}~\bibnamefont{Tsujikawa}},
  \bibinfo{journal}{Living Reviews in Relativity} \textbf{\bibinfo{volume}{13}}
  (\bibinfo{year}{2010}), \urlprefix\url{https://doi.org/10.12942/lrr-2010-3}.

\bibitem[{\citenamefont{Hu and Sawicki}(2007)}]{Hu:2007nk}
\bibinfo{author}{\bibfnamefont{W.}~\bibnamefont{Hu}} \bibnamefont{and}
  \bibinfo{author}{\bibfnamefont{I.}~\bibnamefont{Sawicki}},
  \bibinfo{journal}{Physical Review D} \textbf{\bibinfo{volume}{76}},
  \bibinfo{pages}{064004} (\bibinfo{year}{2007}),
  \urlprefix\url{https://doi.org/10.1103/physrevd.76.064004}.

\bibitem[{\citenamefont{Capozziello et~al.}(2005)\citenamefont{Capozziello,
  Cardone, and Troisi}}]{Capozziello:2005ku}
\bibinfo{author}{\bibfnamefont{S.}~\bibnamefont{Capozziello}},
  \bibinfo{author}{\bibfnamefont{V.~F.} \bibnamefont{Cardone}},
  \bibnamefont{and} \bibinfo{author}{\bibfnamefont{A.}~\bibnamefont{Troisi}},
  \bibinfo{journal}{Physical Review D} \textbf{\bibinfo{volume}{71}},
  \bibinfo{pages}{043503} (\bibinfo{year}{2005}),
  \urlprefix\url{https://doi.org/10.1103/physrevd.71.043503}.

\bibitem[{\citenamefont{Oikonomou}(2022)}]{Oikonomou:2022irx}
\bibinfo{author}{\bibfnamefont{V.}~\bibnamefont{Oikonomou}},
  \bibinfo{journal}{Nuclear Physics B} \textbf{\bibinfo{volume}{984}},
  \bibinfo{pages}{115985} (\bibinfo{year}{2022}),
  \urlprefix\url{https://doi.org/10.1016/j.nuclphysb.2022.115985}.

\bibitem[{\citenamefont{Capozziello et~al.}(2019)\citenamefont{Capozziello,
  D'Agostino, and Luongo}}]{Rocco}
\bibinfo{author}{\bibfnamefont{S.}~\bibnamefont{Capozziello}},
  \bibinfo{author}{\bibfnamefont{R.}~\bibnamefont{D'Agostino}},
  \bibnamefont{and} \bibinfo{author}{\bibfnamefont{O.}~\bibnamefont{Luongo}},
  \bibinfo{journal}{International Journal of Modern Physics D}
  \textbf{\bibinfo{volume}{28}}, \bibinfo{pages}{1930016}
  (\bibinfo{year}{2019}), \eprint{1904.01427},
  \urlprefix\url{https://doi.org/10.1142/S0218271819300167}.

\bibitem[{\citenamefont{Bajardi et~al.}(2022)\citenamefont{Bajardi, D'Agostino,
  Benetti, De~Falco, and Capozziello}}]{Bajardi}
\bibinfo{author}{\bibfnamefont{F.}~\bibnamefont{Bajardi}},
  \bibinfo{author}{\bibfnamefont{R.}~\bibnamefont{D'Agostino}},
  \bibinfo{author}{\bibfnamefont{M.}~\bibnamefont{Benetti}},
  \bibinfo{author}{\bibfnamefont{V.}~\bibnamefont{De~Falco}}, \bibnamefont{and}
  \bibinfo{author}{\bibfnamefont{S.}~\bibnamefont{Capozziello}},
  \bibinfo{journal}{European Physical Journal Plus}
  \textbf{\bibinfo{volume}{137}}, \bibinfo{pages}{1239} (\bibinfo{year}{2022}),
  \eprint{2211.06268},
  \urlprefix\url{https://doi.org/10.1140/epjp/s13360-022-03418-8}.

\bibitem[{\citenamefont{Starobinsky}(1979)}]{Starobinsky1979}
\bibinfo{author}{\bibfnamefont{A.}~\bibnamefont{Starobinsky}},
  \bibinfo{journal}{Journal of Experimental and Theoretical Physics Letters}
  \textbf{\bibinfo{volume}{30}}, \bibinfo{pages}{682} (\bibinfo{year}{1979}).

\bibitem[{\citenamefont{Cai et~al.}(2016)\citenamefont{Cai, Capozziello,
  De~Laurentis, and Saridakis}}]{Cai}
\bibinfo{author}{\bibfnamefont{Y.-F.} \bibnamefont{Cai}},
  \bibinfo{author}{\bibfnamefont{S.}~\bibnamefont{Capozziello}},
  \bibinfo{author}{\bibfnamefont{M.}~\bibnamefont{De~Laurentis}},
  \bibnamefont{and} \bibinfo{author}{\bibfnamefont{E.~N.}
  \bibnamefont{Saridakis}}, \bibinfo{journal}{Reports on Progress in Physics}
  \textbf{\bibinfo{volume}{79}}, \bibinfo{pages}{106901}
  (\bibinfo{year}{2016}),
  \urlprefix\url{https://doi.org/10.1088/0034-4885/79/10/106901}.

\bibitem[{\citenamefont{Nojiri et~al.}(2022)\citenamefont{Nojiri, Odintsov, and
  Oikonomou}}]{Nojiri:2022ski}
\bibinfo{author}{\bibfnamefont{S.}~\bibnamefont{Nojiri}},
  \bibinfo{author}{\bibfnamefont{S.}~\bibnamefont{Odintsov}}, \bibnamefont{and}
  \bibinfo{author}{\bibfnamefont{V.}~\bibnamefont{Oikonomou}},
  \bibinfo{journal}{Nuclear Physics B} \textbf{\bibinfo{volume}{980}},
  \bibinfo{pages}{115850} (\bibinfo{year}{2022}),
  \urlprefix\url{https://doi.org/10.1016/j.nuclphysb.2022.115850}.

\bibitem[{\citenamefont{Spallicci et~al.}(2021)\citenamefont{Spallicci,
  Helay{\"{e}}l-Neto, L{\'{o}}pez-Corredoira, and
  Capozziello}}]{spallicci-etal-2021}
\bibinfo{author}{\bibfnamefont{A.~D. A.~M.} \bibnamefont{Spallicci}},
  \bibinfo{author}{\bibfnamefont{J.~A.} \bibnamefont{Helay{\"{e}}l-Neto}},
  \bibinfo{author}{\bibfnamefont{M.}~\bibnamefont{L{\'{o}}pez-Corredoira}},
  \bibnamefont{and}
  \bibinfo{author}{\bibfnamefont{S.}~\bibnamefont{Capozziello}},
  \bibinfo{journal}{The European Physical Journal C}
  \textbf{\bibinfo{volume}{81}}, \bibinfo{pages}{4} (\bibinfo{year}{2021}),
  \urlprefix\url{https://doi.org/10.1140/epjc/s10052-020-08703-3}.

\bibitem[{\citenamefont{Spallicci
  et~al.}(2022{\natexlab{a}})\citenamefont{Spallicci, Sarracino, and
  Capozziello}}]{Spallicci_2022}
\bibinfo{author}{\bibfnamefont{A.~D. A.~M.} \bibnamefont{Spallicci}},
  \bibinfo{author}{\bibfnamefont{G.}~\bibnamefont{Sarracino}},
  \bibnamefont{and}
  \bibinfo{author}{\bibfnamefont{S.}~\bibnamefont{Capozziello}},
  \bibinfo{journal}{The European Physical Journal Plus}
  \textbf{\bibinfo{volume}{137}}, \bibinfo{pages}{253}
  (\bibinfo{year}{2022}{\natexlab{a}}),
  \urlprefix\url{https://doi.org/10.1140/epjp/s13360-022-02450-y}.

\bibitem[{\citenamefont{Sarracino et~al.}(2022)\citenamefont{Sarracino,
  Spallicci, and Capozziello}}]{Sarracino_2022}
\bibinfo{author}{\bibfnamefont{G.}~\bibnamefont{Sarracino}},
  \bibinfo{author}{\bibfnamefont{A.~D. A.~M.} \bibnamefont{Spallicci}},
  \bibnamefont{and}
  \bibinfo{author}{\bibfnamefont{S.}~\bibnamefont{Capozziello}},
  \bibinfo{journal}{The European Physical Journal Plus}
  \textbf{\bibinfo{volume}{137}}, \bibinfo{pages}{1386} (\bibinfo{year}{2022}),
  \urlprefix\url{https://doi.org/10.1140/epjp/s13360-022-03595-6}.

\bibitem[{\citenamefont{Capozziello et~al.}(2020)\citenamefont{Capozziello,
  Benetti, and Spallicci}}]{capozziello-benetti-spallicci-2020}
\bibinfo{author}{\bibfnamefont{S.}~\bibnamefont{Capozziello}},
  \bibinfo{author}{\bibfnamefont{M.}~\bibnamefont{Benetti}}, \bibnamefont{and}
  \bibinfo{author}{\bibfnamefont{A.~D. A.~M.} \bibnamefont{Spallicci}},
  \bibinfo{journal}{Foundations of Physics} \textbf{\bibinfo{volume}{50}},
  \bibinfo{pages}{893} (\bibinfo{year}{2020}),
  \urlprefix\url{https://doi.org/10.1007/s10701-020-00356-2}.

\bibitem[{\citenamefont{Spallicci
  et~al.}(2022{\natexlab{b}})\citenamefont{Spallicci, Benetti, and
  Capozziello}}]{spallicci-benetti-capozziello-2022}
\bibinfo{author}{\bibfnamefont{A.~D. A.~M.} \bibnamefont{Spallicci}},
  \bibinfo{author}{\bibfnamefont{M.}~\bibnamefont{Benetti}}, \bibnamefont{and}
  \bibinfo{author}{\bibfnamefont{S.}~\bibnamefont{Capozziello}},
  \bibinfo{journal}{Foundations of Physics} \textbf{\bibinfo{volume}{52}},
  \bibinfo{pages}{23} (\bibinfo{year}{2022}{\natexlab{b}}),
  \urlprefix\url{https://doi.org/10.1007/s10701-021-00531-z}.

\bibitem[{\citenamefont{Bernal et~al.}(2016)\citenamefont{Bernal, Verde, and
  Riess}}]{Bernal_2016}
\bibinfo{author}{\bibfnamefont{J.~L.} \bibnamefont{Bernal}},
  \bibinfo{author}{\bibfnamefont{L.}~\bibnamefont{Verde}}, \bibnamefont{and}
  \bibinfo{author}{\bibfnamefont{A.~G.} \bibnamefont{Riess}},
  \bibinfo{journal}{Journal of Cosmology and Astroparticle Physics}
  \textbf{\bibinfo{volume}{2016}}, \bibinfo{pages}{019} (\bibinfo{year}{2016}),
  \urlprefix\url{https://doi.org/10.1088/1475-7516/2016/10/019}.

\bibitem[{\citenamefont{Mörtsell and Dhawan}(2018)}]{Mortsell_2018}
\bibinfo{author}{\bibfnamefont{E.}~\bibnamefont{Mörtsell}} \bibnamefont{and}
  \bibinfo{author}{\bibfnamefont{S.}~\bibnamefont{Dhawan}},
  \bibinfo{journal}{Journal of Cosmology and Astroparticle Physics}
  \textbf{\bibinfo{volume}{2018}}, \bibinfo{pages}{025} (\bibinfo{year}{2018}),
  \urlprefix\url{https://doi.org/10.1088/1475-7516/2018/09/025}.

\bibitem[{\citenamefont{Vagnozzi et~al.}(2018)\citenamefont{Vagnozzi, Dhawan,
  Gerbino, Freese, Goobar, and Mena}}]{Vagnozzi_2018}
\bibinfo{author}{\bibfnamefont{S.}~\bibnamefont{Vagnozzi}},
  \bibinfo{author}{\bibfnamefont{S.}~\bibnamefont{Dhawan}},
  \bibinfo{author}{\bibfnamefont{M.}~\bibnamefont{Gerbino}},
  \bibinfo{author}{\bibfnamefont{K.}~\bibnamefont{Freese}},
  \bibinfo{author}{\bibfnamefont{A.}~\bibnamefont{Goobar}}, \bibnamefont{and}
  \bibinfo{author}{\bibfnamefont{O.}~\bibnamefont{Mena}},
  \bibinfo{journal}{Physical Review D} \textbf{\bibinfo{volume}{98}},
  \bibinfo{pages}{083501} (\bibinfo{year}{2018}),
  \urlprefix\url{https://doi.org/10.1103/physrevd.98.083501}.

\bibitem[{\citenamefont{Yang et~al.}(2018)\citenamefont{Yang, Pan, Valentino,
  Nunes, Vagnozzi, and Mota}}]{Yang_2018}
\bibinfo{author}{\bibfnamefont{W.}~\bibnamefont{Yang}},
  \bibinfo{author}{\bibfnamefont{S.}~\bibnamefont{Pan}},
  \bibinfo{author}{\bibfnamefont{E.~D.} \bibnamefont{Valentino}},
  \bibinfo{author}{\bibfnamefont{R.~C.} \bibnamefont{Nunes}},
  \bibinfo{author}{\bibfnamefont{S.}~\bibnamefont{Vagnozzi}}, \bibnamefont{and}
  \bibinfo{author}{\bibfnamefont{D.~F.} \bibnamefont{Mota}},
  \bibinfo{journal}{Journal of Cosmology and Astroparticle Physics}
  \textbf{\bibinfo{volume}{2018}}, \bibinfo{pages}{019} (\bibinfo{year}{2018}),
  \urlprefix\url{https://doi.org/10.1088/1475-7516/2018/09/019}.

\bibitem[{\citenamefont{Poulin et~al.}(2019)\citenamefont{Poulin, Smith,
  Karwal, and Kamionkowski}}]{Poulin_2019}
\bibinfo{author}{\bibfnamefont{V.}~\bibnamefont{Poulin}},
  \bibinfo{author}{\bibfnamefont{T.~L.} \bibnamefont{Smith}},
  \bibinfo{author}{\bibfnamefont{T.}~\bibnamefont{Karwal}}, \bibnamefont{and}
  \bibinfo{author}{\bibfnamefont{M.}~\bibnamefont{Kamionkowski}},
  \bibinfo{journal}{Physical Review Letters} \textbf{\bibinfo{volume}{122}},
  \bibinfo{pages}{221301} (\bibinfo{year}{2019}),
  \urlprefix\url{https://doi.org/10.1103/physrevlett.122.221301}.

\bibitem[{\citenamefont{Kreisch et~al.}(2020)\citenamefont{Kreisch, Cyr-Racine,
  and Dor{\'{e}}}}]{Kreisch_2020}
\bibinfo{author}{\bibfnamefont{C.~D.} \bibnamefont{Kreisch}},
  \bibinfo{author}{\bibfnamefont{F.-Y.} \bibnamefont{Cyr-Racine}},
  \bibnamefont{and}
  \bibinfo{author}{\bibfnamefont{O.}~\bibnamefont{Dor{\'{e}}}},
  \bibinfo{journal}{Physical Review D} \textbf{\bibinfo{volume}{101}},
  \bibinfo{pages}{103520} (\bibinfo{year}{2020}),
  \urlprefix\url{https://doi.org/10.1103/physrevd.101.123505}.

\bibitem[{\citenamefont{Agrawal et~al.}(2019)\citenamefont{Agrawal, Cyr-Racine,
  Pinner, and Randall}}]{Agrawal_2019}
\bibinfo{author}{\bibfnamefont{P.}~\bibnamefont{Agrawal}},
  \bibinfo{author}{\bibfnamefont{F.-Y.} \bibnamefont{Cyr-Racine}},
  \bibinfo{author}{\bibfnamefont{D.}~\bibnamefont{Pinner}}, \bibnamefont{and}
  \bibinfo{author}{\bibfnamefont{L.}~\bibnamefont{Randall}}
  (\bibinfo{year}{2019}), \urlprefix\url{https://arxiv.org/abs/1904.01016}.

\bibitem[{\citenamefont{Valentino et~al.}(2019)\citenamefont{Valentino,
  Ferreira, Visinelli, and Danielsson}}]{Di_Valentino_2019}
\bibinfo{author}{\bibfnamefont{E.~D.} \bibnamefont{Valentino}},
  \bibinfo{author}{\bibfnamefont{R.~Z.} \bibnamefont{Ferreira}},
  \bibinfo{author}{\bibfnamefont{L.}~\bibnamefont{Visinelli}},
  \bibnamefont{and}
  \bibinfo{author}{\bibfnamefont{U.}~\bibnamefont{Danielsson}},
  \bibinfo{journal}{Physics of the Dark Universe}
  \textbf{\bibinfo{volume}{26}}, \bibinfo{pages}{100385}
  (\bibinfo{year}{2019}),
  \urlprefix\url{https://doi.org/10.1016/j.dark.2019.100385}.

\bibitem[{\citenamefont{Pan et~al.}(2019)\citenamefont{Pan, Yang, Valentino,
  Saridakis, and Chakraborty}}]{Pan_2019}
\bibinfo{author}{\bibfnamefont{S.}~\bibnamefont{Pan}},
  \bibinfo{author}{\bibfnamefont{W.}~\bibnamefont{Yang}},
  \bibinfo{author}{\bibfnamefont{E.~D.} \bibnamefont{Valentino}},
  \bibinfo{author}{\bibfnamefont{E.~N.} \bibnamefont{Saridakis}},
  \bibnamefont{and}
  \bibinfo{author}{\bibfnamefont{S.}~\bibnamefont{Chakraborty}},
  \bibinfo{journal}{Physical Review D} \textbf{\bibinfo{volume}{100}},
  \bibinfo{pages}{103520} (\bibinfo{year}{2019}),
  \urlprefix\url{https://doi.org/10.1103/physrevd.100.103520}.

\bibitem[{\citenamefont{Vagnozzi}(2020)}]{Vagnozzi_2020}
\bibinfo{author}{\bibfnamefont{S.}~\bibnamefont{Vagnozzi}},
  \bibinfo{journal}{Physical Review D} \textbf{\bibinfo{volume}{102}},
  \bibinfo{pages}{023518} (\bibinfo{year}{2020}),
  \urlprefix\url{https://doi.org/10.1103/physrevd.102.023518}.

\bibitem[{\citenamefont{Visinelli et~al.}(2019)\citenamefont{Visinelli,
  Vagnozzi, and Danielsson}}]{Visinelli_2019}
\bibinfo{author}{\bibfnamefont{L.}~\bibnamefont{Visinelli}},
  \bibinfo{author}{\bibfnamefont{S.}~\bibnamefont{Vagnozzi}}, \bibnamefont{and}
  \bibinfo{author}{\bibfnamefont{U.}~\bibnamefont{Danielsson}},
  \bibinfo{journal}{Symmetry} \textbf{\bibinfo{volume}{11}},
  \bibinfo{pages}{1035} (\bibinfo{year}{2019}),
  \urlprefix\url{https://doi.org/10.3390/sym11081035}.

\bibitem[{\citenamefont{Knox and Millea}(2020)}]{Knox_2020}
\bibinfo{author}{\bibfnamefont{L.}~\bibnamefont{Knox}} \bibnamefont{and}
  \bibinfo{author}{\bibfnamefont{M.}~\bibnamefont{Millea}},
  \bibinfo{journal}{Physical Review D} \textbf{\bibinfo{volume}{101}},
  \bibinfo{pages}{043533} (\bibinfo{year}{2020}),
  \urlprefix\url{https://doi.org/10.1103/physrevd.101.043533}.

\bibitem[{\citenamefont{Valentino
  et~al.}(2020{\natexlab{a}})\citenamefont{Valentino, Melchiorri, Mena, and
  Vagnozzi}}]{Di_Valentino_2020}
\bibinfo{author}{\bibfnamefont{E.~D.} \bibnamefont{Valentino}},
  \bibinfo{author}{\bibfnamefont{A.}~\bibnamefont{Melchiorri}},
  \bibinfo{author}{\bibfnamefont{O.}~\bibnamefont{Mena}}, \bibnamefont{and}
  \bibinfo{author}{\bibfnamefont{S.}~\bibnamefont{Vagnozzi}},
  \bibinfo{journal}{Physics of the Dark Universe}
  \textbf{\bibinfo{volume}{30}}, \bibinfo{pages}{100666}
  (\bibinfo{year}{2020}{\natexlab{a}}),
  \urlprefix\url{https://doi.org/10.1016/j.dark.2020.100666}.

\bibitem[{\citenamefont{Valentino
  et~al.}(2020{\natexlab{b}})\citenamefont{Valentino, Melchiorri, Mena, and
  Vagnozzi}}]{Di_Valentino_2020b}
\bibinfo{author}{\bibfnamefont{E.~D.} \bibnamefont{Valentino}},
  \bibinfo{author}{\bibfnamefont{A.}~\bibnamefont{Melchiorri}},
  \bibinfo{author}{\bibfnamefont{O.}~\bibnamefont{Mena}}, \bibnamefont{and}
  \bibinfo{author}{\bibfnamefont{S.}~\bibnamefont{Vagnozzi}},
  \bibinfo{journal}{Physical Review D} \textbf{\bibinfo{volume}{101}},
  \bibinfo{pages}{063502} (\bibinfo{year}{2020}{\natexlab{b}}),
  \urlprefix\url{https://doi.org/10.1103/physrevd.101.063502}.

\bibitem[{\citenamefont{Valentino et~al.}(2021)\citenamefont{Valentino,
  Mukherjee, and Sen}}]{Di_Valentino_2021f}
\bibinfo{author}{\bibfnamefont{E.~D.} \bibnamefont{Valentino}},
  \bibinfo{author}{\bibfnamefont{A.}~\bibnamefont{Mukherjee}},
  \bibnamefont{and} \bibinfo{author}{\bibfnamefont{A.~A.} \bibnamefont{Sen}},
  \bibinfo{journal}{Entropy} \textbf{\bibinfo{volume}{23}},
  \bibinfo{pages}{404} (\bibinfo{year}{2021}),
  \urlprefix\url{https://doi.org/10.3390/e23040404}.

\bibitem[{\citenamefont{Smith et~al.}(2021)\citenamefont{Smith, Poulin, Bernal,
  Boddy, Kamionkowski, and Murgia}}]{Smith_2021}
\bibinfo{author}{\bibfnamefont{T.~L.} \bibnamefont{Smith}},
  \bibinfo{author}{\bibfnamefont{V.}~\bibnamefont{Poulin}},
  \bibinfo{author}{\bibfnamefont{J.~L.} \bibnamefont{Bernal}},
  \bibinfo{author}{\bibfnamefont{K.~K.} \bibnamefont{Boddy}},
  \bibinfo{author}{\bibfnamefont{M.}~\bibnamefont{Kamionkowski}},
  \bibnamefont{and} \bibinfo{author}{\bibfnamefont{R.}~\bibnamefont{Murgia}},
  \bibinfo{journal}{Physical Review D} \textbf{\bibinfo{volume}{103}},
  \bibinfo{pages}{123542} (\bibinfo{year}{2021}),
  \urlprefix\url{https://doi.org/10.1103/physrevd.103.123542}.

\bibitem[{\citenamefont{Vagnozzi}(2021)}]{Vagnozzi_2021}
\bibinfo{author}{\bibfnamefont{S.}~\bibnamefont{Vagnozzi}},
  \bibinfo{journal}{Physical Review D} \textbf{\bibinfo{volume}{104}},
  \bibinfo{pages}{063524} (\bibinfo{year}{2021}),
  \urlprefix\url{https://doi.org/10.1103/physrevd.104.063524}.

\bibitem[{\citenamefont{Nunes and Valentino}(2021)}]{Nunes_2021}
\bibinfo{author}{\bibfnamefont{R.~C.} \bibnamefont{Nunes}} \bibnamefont{and}
  \bibinfo{author}{\bibfnamefont{E.~D.} \bibnamefont{Valentino}},
  \bibinfo{journal}{Physical Review D} \textbf{\bibinfo{volume}{104}},
  \bibinfo{pages}{063529} (\bibinfo{year}{2021}),
  \urlprefix\url{https://doi.org/10.1103/physrevd.104.063529}.

\bibitem[{\citenamefont{Cyr-Racine et~al.}(2022)\citenamefont{Cyr-Racine, Ge,
  and Knox}}]{Cyr_Racine_2022}
\bibinfo{author}{\bibfnamefont{F.-Y.} \bibnamefont{Cyr-Racine}},
  \bibinfo{author}{\bibfnamefont{F.}~\bibnamefont{Ge}}, \bibnamefont{and}
  \bibinfo{author}{\bibfnamefont{L.}~\bibnamefont{Knox}},
  \bibinfo{journal}{Physical Review Letters} \textbf{\bibinfo{volume}{128}},
  \bibinfo{pages}{201301} (\bibinfo{year}{2022}),
  \urlprefix\url{https://doi.org/10.1103/physrevlett.128.201301}.

\bibitem[{\citenamefont{Anchordoqui et~al.}(2021)\citenamefont{Anchordoqui,
  Valentino, Pan, and Yang}}]{Anchordoqui_2021}
\bibinfo{author}{\bibfnamefont{L.~A.} \bibnamefont{Anchordoqui}},
  \bibinfo{author}{\bibfnamefont{E.~D.} \bibnamefont{Valentino}},
  \bibinfo{author}{\bibfnamefont{S.}~\bibnamefont{Pan}}, \bibnamefont{and}
  \bibinfo{author}{\bibfnamefont{W.}~\bibnamefont{Yang}},
  \bibinfo{journal}{Journal of High Energy Astrophysics}
  \textbf{\bibinfo{volume}{32}}, \bibinfo{pages}{28} (\bibinfo{year}{2021}),
  \urlprefix\url{https://doi.org/10.1016/j.jheap.2021.08.001}.

\bibitem[{\citenamefont{Poulin et~al.}(2021)\citenamefont{Poulin, Smith, and
  Bartlett}}]{Poulin_2021}
\bibinfo{author}{\bibfnamefont{V.}~\bibnamefont{Poulin}},
  \bibinfo{author}{\bibfnamefont{T.~L.} \bibnamefont{Smith}}, \bibnamefont{and}
  \bibinfo{author}{\bibfnamefont{A.}~\bibnamefont{Bartlett}},
  \bibinfo{journal}{Physical Review D} \textbf{\bibinfo{volume}{104}},
  \bibinfo{pages}{123550} (\bibinfo{year}{2021}),
  \urlprefix\url{https://doi.org/10.1103/physrevd.104.123550}.

\bibitem[{\citenamefont{Alestas et~al.}(2022)\citenamefont{Alestas, Camarena,
  Valentino, Kazantzidis, Marra, Nesseris, and
  Perivolaropoulos}}]{Alestas_2022}
\bibinfo{author}{\bibfnamefont{G.}~\bibnamefont{Alestas}},
  \bibinfo{author}{\bibfnamefont{D.}~\bibnamefont{Camarena}},
  \bibinfo{author}{\bibfnamefont{E.~D.} \bibnamefont{Valentino}},
  \bibinfo{author}{\bibfnamefont{L.}~\bibnamefont{Kazantzidis}},
  \bibinfo{author}{\bibfnamefont{V.}~\bibnamefont{Marra}},
  \bibinfo{author}{\bibfnamefont{S.}~\bibnamefont{Nesseris}}, \bibnamefont{and}
  \bibinfo{author}{\bibfnamefont{L.}~\bibnamefont{Perivolaropoulos}},
  \bibinfo{journal}{Physical Review D} \textbf{\bibinfo{volume}{105}},
  \bibinfo{pages}{063538} (\bibinfo{year}{2022}),
  \urlprefix\url{https://doi.org/10.1103/physrevd.105.063538}.

\bibitem[{\citenamefont{Smith et~al.}(2022)\citenamefont{Smith, Lucca, Poulin,
  Abellan, Balkenhol, Benabed, Galli, and Murgia}}]{Smith_2022}
\bibinfo{author}{\bibfnamefont{T.~L.} \bibnamefont{Smith}},
  \bibinfo{author}{\bibfnamefont{M.}~\bibnamefont{Lucca}},
  \bibinfo{author}{\bibfnamefont{V.}~\bibnamefont{Poulin}},
  \bibinfo{author}{\bibfnamefont{G.~F.} \bibnamefont{Abellan}},
  \bibinfo{author}{\bibfnamefont{L.}~\bibnamefont{Balkenhol}},
  \bibinfo{author}{\bibfnamefont{K.}~\bibnamefont{Benabed}},
  \bibinfo{author}{\bibfnamefont{S.}~\bibnamefont{Galli}}, \bibnamefont{and}
  \bibinfo{author}{\bibfnamefont{R.}~\bibnamefont{Murgia}},
  \bibinfo{journal}{Physical Review D} \textbf{\bibinfo{volume}{106}},
  \bibinfo{pages}{043526} (\bibinfo{year}{2022}),
  \urlprefix\url{https://doi.org/10.1103/physrevd.106.043526}.

\bibitem[{\citenamefont{Reeves et~al.}(2023)\citenamefont{Reeves, Herold,
  Vagnozzi, Sherwin, and Ferreira}}]{Reeves_2023}
\bibinfo{author}{\bibfnamefont{A.}~\bibnamefont{Reeves}},
  \bibinfo{author}{\bibfnamefont{L.}~\bibnamefont{Herold}},
  \bibinfo{author}{\bibfnamefont{S.}~\bibnamefont{Vagnozzi}},
  \bibinfo{author}{\bibfnamefont{B.~D.} \bibnamefont{Sherwin}},
  \bibnamefont{and} \bibinfo{author}{\bibfnamefont{E.~G.~M.}
  \bibnamefont{Ferreira}}, \bibinfo{journal}{Monthly Notices of the Royal
  Astronomical Society} \textbf{\bibinfo{volume}{520}}, \bibinfo{pages}{3688}
  (\bibinfo{year}{2023}),
  \urlprefix\url{https://doi.org/10.1093/mnras/stad317}.

\bibitem[{\citenamefont{Poulin et~al.}(2023)\citenamefont{Poulin, Smith, and
  Karwal}}]{Poulin_2023}
\bibinfo{author}{\bibfnamefont{V.}~\bibnamefont{Poulin}},
  \bibinfo{author}{\bibfnamefont{T.~L.} \bibnamefont{Smith}}, \bibnamefont{and}
  \bibinfo{author}{\bibfnamefont{T.}~\bibnamefont{Karwal}}
  (\bibinfo{year}{2023}), \urlprefix\url{https://arxiv.org/abs/2302.09032}.

\bibitem[{\citenamefont{Demianski et~al.}(2017)\citenamefont{Demianski,
  Piedipalumbo, Sawant, and Amati}}]{Ester1}
\bibinfo{author}{\bibfnamefont{M.}~\bibnamefont{Demianski}},
  \bibinfo{author}{\bibfnamefont{E.}~\bibnamefont{Piedipalumbo}},
  \bibinfo{author}{\bibfnamefont{D.}~\bibnamefont{Sawant}}, \bibnamefont{and}
  \bibinfo{author}{\bibfnamefont{L.}~\bibnamefont{Amati}},
  \bibinfo{journal}{Astronomy \& Astrophysics} \textbf{\bibinfo{volume}{598}},
  \bibinfo{pages}{A112} (\bibinfo{year}{2017}),
  \urlprefix\url{https://doi.org/10.1051/0004-6361/201628909}.

\bibitem[{\citenamefont{Lusso et~al.}(2019)\citenamefont{Lusso, Piedipalumbo,
  Risaliti, Paolillo, Bisogni, Nardini, and Amati}}]{Ester2}
\bibinfo{author}{\bibfnamefont{E.}~\bibnamefont{Lusso}},
  \bibinfo{author}{\bibfnamefont{E.}~\bibnamefont{Piedipalumbo}},
  \bibinfo{author}{\bibfnamefont{G.}~\bibnamefont{Risaliti}},
  \bibinfo{author}{\bibfnamefont{M.}~\bibnamefont{Paolillo}},
  \bibinfo{author}{\bibfnamefont{S.}~\bibnamefont{Bisogni}},
  \bibinfo{author}{\bibfnamefont{E.}~\bibnamefont{Nardini}}, \bibnamefont{and}
  \bibinfo{author}{\bibfnamefont{L.}~\bibnamefont{Amati}},
  \bibinfo{journal}{Astronomy \& Astrophysics} \textbf{\bibinfo{volume}{628}},
  \bibinfo{pages}{L4} (\bibinfo{year}{2019}),
  \urlprefix\url{https://doi.org/10.1051/0004-6361/201936223}.

\bibitem[{\citenamefont{Ryan et~al.}(2019)\citenamefont{Ryan, Chen, and
  Ratra}}]{Ratra_1}
\bibinfo{author}{\bibfnamefont{J.}~\bibnamefont{Ryan}},
  \bibinfo{author}{\bibfnamefont{Y.}~\bibnamefont{Chen}}, \bibnamefont{and}
  \bibinfo{author}{\bibfnamefont{B.}~\bibnamefont{Ratra}},
  \bibinfo{journal}{Monthly Notices of the Royal Astronomical Society}
  \textbf{\bibinfo{volume}{488}}, \bibinfo{pages}{3844} (\bibinfo{year}{2019}),
  \urlprefix\url{https://doi.org/10.1093/mnras/stz1966}.

\bibitem[{\citenamefont{Cao et~al.}(2021)\citenamefont{Cao, Ryan, and
  Ratra}}]{Ratra_2}
\bibinfo{author}{\bibfnamefont{S.}~\bibnamefont{Cao}},
  \bibinfo{author}{\bibfnamefont{J.}~\bibnamefont{Ryan}}, \bibnamefont{and}
  \bibinfo{author}{\bibfnamefont{B.}~\bibnamefont{Ratra}},
  \bibinfo{journal}{Monthly Notices of the Royal Astronomical Society}
  \textbf{\bibinfo{volume}{504}}, \bibinfo{pages}{300} (\bibinfo{year}{2021}),
  \urlprefix\url{https://doi.org/10.1093/mnras/stab942}.

\bibitem[{\citenamefont{Krishnan
  et~al.}(2021{\natexlab{a}})\citenamefont{Krishnan, Colg{\'{a}}in,
  Sheikh-Jabbari, and Yang}}]{Krishnan_2021}
\bibinfo{author}{\bibfnamefont{C.}~\bibnamefont{Krishnan}},
  \bibinfo{author}{\bibfnamefont{E.~{\'{O} }.} \bibnamefont{Colg{\'{a}}in}},
  \bibinfo{author}{\bibfnamefont{M.}~\bibnamefont{Sheikh-Jabbari}},
  \bibnamefont{and} \bibinfo{author}{\bibfnamefont{T.}~\bibnamefont{Yang}},
  \bibinfo{journal}{Physical Review D} \textbf{\bibinfo{volume}{103}},
  \bibinfo{pages}{103509} (\bibinfo{year}{2021}{\natexlab{a}}),
  \urlprefix\url{https://doi.org/10.1103/physrevd.103.103509}.

\bibitem[{\citenamefont{Colg{\'{a}}in et~al.}(2021)\citenamefont{Colg{\'{a}}in,
  Sheikh-Jabbari, and Yang}}]{Colgain_2021}
\bibinfo{author}{\bibfnamefont{E.~{\'{O} }.} \bibnamefont{Colg{\'{a}}in}},
  \bibinfo{author}{\bibfnamefont{M.}~\bibnamefont{Sheikh-Jabbari}},
  \bibnamefont{and} \bibinfo{author}{\bibfnamefont{T.}~\bibnamefont{Yang}},
  \bibinfo{journal}{The European Physical Journal C}
  \textbf{\bibinfo{volume}{81}}, \bibinfo{pages}{892} (\bibinfo{year}{2021}),
  \urlprefix\url{https://doi.org/10.1140/epjc/s10052-021-09708-2}.

\bibitem[{\citenamefont{Krishnan and Mondol}(2022)}]{Krishnan_2022}
\bibinfo{author}{\bibfnamefont{C.}~\bibnamefont{Krishnan}} \bibnamefont{and}
  \bibinfo{author}{\bibfnamefont{R.}~\bibnamefont{Mondol}}
  (\bibinfo{year}{2022}), \urlprefix\url{https://arxiv.org/abs/2201.13384}.

\bibitem[{\citenamefont{Colg{\'{a}}in
  et~al.}(2022{\natexlab{a}})\citenamefont{Colg{\'{a}}in, Sheikh-Jabbari,
  Solomon, Bargiacchi, Capozziello, Dainotti, and Stojkovic}}]{Colgain_2022}
\bibinfo{author}{\bibfnamefont{E.~{\'{O} }.} \bibnamefont{Colg{\'{a}}in}},
  \bibinfo{author}{\bibfnamefont{M.}~\bibnamefont{Sheikh-Jabbari}},
  \bibinfo{author}{\bibfnamefont{R.}~\bibnamefont{Solomon}},
  \bibinfo{author}{\bibfnamefont{G.}~\bibnamefont{Bargiacchi}},
  \bibinfo{author}{\bibfnamefont{S.}~\bibnamefont{Capozziello}},
  \bibinfo{author}{\bibfnamefont{M.}~\bibnamefont{Dainotti}}, \bibnamefont{and}
  \bibinfo{author}{\bibfnamefont{D.}~\bibnamefont{Stojkovic}},
  \bibinfo{journal}{Physical Review D} \textbf{\bibinfo{volume}{106}},
  \bibinfo{pages}{L041301} (\bibinfo{year}{2022}{\natexlab{a}}),
  \urlprefix\url{https://doi.org/10.1103/physrevd.106.l041301}.

\bibitem[{\citenamefont{Colg{\'{a}}in
  et~al.}(2022{\natexlab{b}})\citenamefont{Colg{\'{a}}in, Sheikh-Jabbari,
  Solomon, Dainotti, and Stojkovic}}]{Colgain_2022b}
\bibinfo{author}{\bibfnamefont{E.~{\'{O} }.} \bibnamefont{Colg{\'{a}}in}},
  \bibinfo{author}{\bibfnamefont{M.~M.} \bibnamefont{Sheikh-Jabbari}},
  \bibinfo{author}{\bibfnamefont{R.}~\bibnamefont{Solomon}},
  \bibinfo{author}{\bibfnamefont{M.~G.} \bibnamefont{Dainotti}},
  \bibnamefont{and} \bibinfo{author}{\bibfnamefont{D.}~\bibnamefont{Stojkovic}}
  (\bibinfo{year}{2022}{\natexlab{b}}),
  \urlprefix\url{https://arxiv.org/abs/2206.11447}.

\bibitem[{\citenamefont{Colg{\'{a}}in
  et~al.}(2022{\natexlab{c}})\citenamefont{Colg{\'{a}}in, Sheikh-Jabbari, and
  Solomon}}]{Colgain_2022c}
\bibinfo{author}{\bibfnamefont{E.~{\'{O} }.} \bibnamefont{Colg{\'{a}}in}},
  \bibinfo{author}{\bibfnamefont{M.~M.} \bibnamefont{Sheikh-Jabbari}},
  \bibnamefont{and} \bibinfo{author}{\bibfnamefont{R.}~\bibnamefont{Solomon}}
  (\bibinfo{year}{2022}{\natexlab{c}}),
  \urlprefix\url{https://arxiv.org/abs/2211.02129}.

\bibitem[{\citenamefont{Dainotti et~al.}(2021)\citenamefont{Dainotti, Simone,
  Schiavone, Montani, Rinaldi, and Lambiase}}]{dainotti_2021a}
\bibinfo{author}{\bibfnamefont{M.~G.} \bibnamefont{Dainotti}},
  \bibinfo{author}{\bibfnamefont{B.~D.} \bibnamefont{Simone}},
  \bibinfo{author}{\bibfnamefont{T.}~\bibnamefont{Schiavone}},
  \bibinfo{author}{\bibfnamefont{G.}~\bibnamefont{Montani}},
  \bibinfo{author}{\bibfnamefont{E.}~\bibnamefont{Rinaldi}}, \bibnamefont{and}
  \bibinfo{author}{\bibfnamefont{G.}~\bibnamefont{Lambiase}},
  \bibinfo{journal}{The Astrophysical Journal} \textbf{\bibinfo{volume}{912}},
  \bibinfo{pages}{150} (\bibinfo{year}{2021}),
  \urlprefix\url{https://doi.org/10.3847/1538-4357/abeb73}.

\bibitem[{\citenamefont{Dainotti
  et~al.}(2022{\natexlab{a}})\citenamefont{Dainotti, Simone, Schiavone,
  Montani, Rinaldi, Lambiase, Bogdan, and Ugale}}]{Dainotti_2022a}
\bibinfo{author}{\bibfnamefont{M.~G.} \bibnamefont{Dainotti}},
  \bibinfo{author}{\bibfnamefont{B.~D.} \bibnamefont{Simone}},
  \bibinfo{author}{\bibfnamefont{T.}~\bibnamefont{Schiavone}},
  \bibinfo{author}{\bibfnamefont{G.}~\bibnamefont{Montani}},
  \bibinfo{author}{\bibfnamefont{E.}~\bibnamefont{Rinaldi}},
  \bibinfo{author}{\bibfnamefont{G.}~\bibnamefont{Lambiase}},
  \bibinfo{author}{\bibfnamefont{M.}~\bibnamefont{Bogdan}}, \bibnamefont{and}
  \bibinfo{author}{\bibfnamefont{S.}~\bibnamefont{Ugale}},
  \bibinfo{journal}{Galaxies} \textbf{\bibinfo{volume}{10}},
  \bibinfo{pages}{24} (\bibinfo{year}{2022}{\natexlab{a}}),
  \urlprefix\url{https://doi.org/10.3390/galaxies10010024}.

\bibitem[{\citenamefont{Dainotti et~al.}(2023)\citenamefont{Dainotti,
  De~Simone, Montani, Schiavone, and Lambiase}}]{Dainotti_2023}
\bibinfo{author}{\bibfnamefont{M.}~\bibnamefont{Dainotti}},
  \bibinfo{author}{\bibfnamefont{B.}~\bibnamefont{De~Simone}},
  \bibinfo{author}{\bibfnamefont{G.}~\bibnamefont{Montani}},
  \bibinfo{author}{\bibfnamefont{T.}~\bibnamefont{Schiavone}},
  \bibnamefont{and} \bibinfo{author}{\bibfnamefont{G.}~\bibnamefont{Lambiase}}
  (\bibinfo{year}{2023}), \urlprefix\url{https://arxiv.org/abs/2301.10572}.

\bibitem[{\citenamefont{Schiavone et~al.}(2022)\citenamefont{Schiavone,
  Montani, Dainotti, De~Simone, Rinaldi, and Lambiase}}]{Schiavone_2022}
\bibinfo{author}{\bibfnamefont{T.}~\bibnamefont{Schiavone}},
  \bibinfo{author}{\bibfnamefont{G.}~\bibnamefont{Montani}},
  \bibinfo{author}{\bibfnamefont{M.~G.} \bibnamefont{Dainotti}},
  \bibinfo{author}{\bibfnamefont{B.}~\bibnamefont{De~Simone}},
  \bibinfo{author}{\bibfnamefont{E.}~\bibnamefont{Rinaldi}}, \bibnamefont{and}
  \bibinfo{author}{\bibfnamefont{G.}~\bibnamefont{Lambiase}}
  (\bibinfo{year}{2022}), \urlprefix\url{https://arxiv.org/abs/2205.07033}.

\bibitem[{\citenamefont{Schiavone and Montani}(2022)}]{Schiavone_2022b}
\bibinfo{author}{\bibfnamefont{T.}~\bibnamefont{Schiavone}} \bibnamefont{and}
  \bibinfo{author}{\bibfnamefont{G.}~\bibnamefont{Montani}}
  (\bibinfo{year}{2022}), \urlprefix\url{https://arxiv.org/abs/2211.16737}.

\bibitem[{\citenamefont{Malekjani et~al.}(2023)\citenamefont{Malekjani,
  Conville, Colg{\'{a}}in, Pourojaghi, and Sheikh-Jabbari}}]{Malekjani_2023}
\bibinfo{author}{\bibfnamefont{M.}~\bibnamefont{Malekjani}},
  \bibinfo{author}{\bibfnamefont{R.~M.} \bibnamefont{Conville}},
  \bibinfo{author}{\bibfnamefont{E.~{\'{O}}.} \bibnamefont{Colg{\'{a}}in}},
  \bibinfo{author}{\bibfnamefont{S.}~\bibnamefont{Pourojaghi}},
  \bibnamefont{and} \bibinfo{author}{\bibfnamefont{M.~M.}
  \bibnamefont{Sheikh-Jabbari}} (\bibinfo{year}{2023}),
  \urlprefix\url{https://arxiv.org/abs/2301.12725}.

\bibitem[{\citenamefont{Gurzadyan et~al.}(2022)\citenamefont{Gurzadyan, Fimin,
  and Chechetkin}}]{Vahe}
\bibinfo{author}{\bibfnamefont{V.~G.} \bibnamefont{Gurzadyan}},
  \bibinfo{author}{\bibfnamefont{N.~N.} \bibnamefont{Fimin}}, \bibnamefont{and}
  \bibinfo{author}{\bibfnamefont{V.~M.} \bibnamefont{Chechetkin}},
  \bibinfo{journal}{Astronomy \& Astrophysics} \textbf{\bibinfo{volume}{666}},
  \bibinfo{pages}{A149} (\bibinfo{year}{2022}), \eprint{2209.00184},
  \urlprefix\url{https://doi.org/10.1051/0004-6361/202244668}.

\bibitem[{\citenamefont{Hu and Wang}(2023)}]{Hu_2023}
\bibinfo{author}{\bibfnamefont{J.-P.} \bibnamefont{Hu}} \bibnamefont{and}
  \bibinfo{author}{\bibfnamefont{F.-Y.} \bibnamefont{Wang}}
  (\bibinfo{year}{2023}), \urlprefix\url{https://arxiv.org/abs/2302.05709}.

\bibitem[{\citenamefont{Jimenez et~al.}(2019)\citenamefont{Jimenez, Cimatti,
  Verde, Moresco, and Wandelt}}]{Jimenez_2019}
\bibinfo{author}{\bibfnamefont{R.}~\bibnamefont{Jimenez}},
  \bibinfo{author}{\bibfnamefont{A.}~\bibnamefont{Cimatti}},
  \bibinfo{author}{\bibfnamefont{L.}~\bibnamefont{Verde}},
  \bibinfo{author}{\bibfnamefont{M.}~\bibnamefont{Moresco}}, \bibnamefont{and}
  \bibinfo{author}{\bibfnamefont{B.}~\bibnamefont{Wandelt}},
  \bibinfo{journal}{Journal of Cosmology and Astroparticle Physics}
  \textbf{\bibinfo{volume}{2019}}, \bibinfo{pages}{043} (\bibinfo{year}{2019}),
  \urlprefix\url{https://doi.org/10.1088/1475-7516/2019/03/043}.

\bibitem[{\citenamefont{Bernal et~al.}(2021)\citenamefont{Bernal, Verde,
  Jimenez, Kamionkowski, Valcin, and Wandelt}}]{Bernal_2021}
\bibinfo{author}{\bibfnamefont{J.~L.} \bibnamefont{Bernal}},
  \bibinfo{author}{\bibfnamefont{L.}~\bibnamefont{Verde}},
  \bibinfo{author}{\bibfnamefont{R.}~\bibnamefont{Jimenez}},
  \bibinfo{author}{\bibfnamefont{M.}~\bibnamefont{Kamionkowski}},
  \bibinfo{author}{\bibfnamefont{D.}~\bibnamefont{Valcin}}, \bibnamefont{and}
  \bibinfo{author}{\bibfnamefont{B.~D.} \bibnamefont{Wandelt}},
  \bibinfo{journal}{Physical Review D} \textbf{\bibinfo{volume}{103}},
  \bibinfo{pages}{103533} (\bibinfo{year}{2021}),
  \urlprefix\url{https://doi.org/10.1103/physrevd.103.103533}.

\bibitem[{\citenamefont{Boylan-Kolchin and Weisz}(2021)}]{Boylan_Kolchin_2021}
\bibinfo{author}{\bibfnamefont{M.}~\bibnamefont{Boylan-Kolchin}}
  \bibnamefont{and} \bibinfo{author}{\bibfnamefont{D.~R.} \bibnamefont{Weisz}},
  \bibinfo{journal}{Monthly Notices of the Royal Astronomical Society}
  \textbf{\bibinfo{volume}{505}}, \bibinfo{pages}{2764} (\bibinfo{year}{2021}),
  \urlprefix\url{https://doi.org/10.1093/mnras/stab1521}.

\bibitem[{\citenamefont{Krishnan
  et~al.}(2021{\natexlab{b}})\citenamefont{Krishnan, Mohayaee, Colg{\'{a}}in,
  Sheikh-Jabbari, and Yin}}]{Krishnan_2021b}
\bibinfo{author}{\bibfnamefont{C.}~\bibnamefont{Krishnan}},
  \bibinfo{author}{\bibfnamefont{R.}~\bibnamefont{Mohayaee}},
  \bibinfo{author}{\bibfnamefont{E.~{\'{O} }.} \bibnamefont{Colg{\'{a}}in}},
  \bibinfo{author}{\bibfnamefont{M.~M.} \bibnamefont{Sheikh-Jabbari}},
  \bibnamefont{and} \bibinfo{author}{\bibfnamefont{L.}~\bibnamefont{Yin}},
  \bibinfo{journal}{Classical and Quantum Gravity}
  \textbf{\bibinfo{volume}{38}}, \bibinfo{pages}{184001}
  (\bibinfo{year}{2021}{\natexlab{b}}),
  \urlprefix\url{https://doi.org/10.1088/1361-6382/ac1a81}.

\bibitem[{\citenamefont{Vagnozzi et~al.}(2022)\citenamefont{Vagnozzi, Pacucci,
  and Loeb}}]{Vagnozzi_2022}
\bibinfo{author}{\bibfnamefont{S.}~\bibnamefont{Vagnozzi}},
  \bibinfo{author}{\bibfnamefont{F.}~\bibnamefont{Pacucci}}, \bibnamefont{and}
  \bibinfo{author}{\bibfnamefont{A.}~\bibnamefont{Loeb}},
  \bibinfo{journal}{Journal of High Energy Astrophysics}
  \textbf{\bibinfo{volume}{36}}, \bibinfo{pages}{27} (\bibinfo{year}{2022}),
  \urlprefix\url{https://doi.org/10.1016/j.jheap.2022.07.004}.

\bibitem[{\citenamefont{Cimatti and Moresco}(2023)}]{Cimatti_2023}
\bibinfo{author}{\bibfnamefont{A.}~\bibnamefont{Cimatti}} \bibnamefont{and}
  \bibinfo{author}{\bibfnamefont{M.}~\bibnamefont{Moresco}}
  (\bibinfo{year}{2023}), \urlprefix\url{https://arxiv.org/abs/2302.07899}.

\bibitem[{\citenamefont{Freedman et~al.}(2019)\citenamefont{Freedman, Madore,
  Hatt, Hoyt, Jang, Beaton, Burns, Lee, Monson, Neeley et~al.}}]{Freedman_2019}
\bibinfo{author}{\bibfnamefont{W.~L.} \bibnamefont{Freedman}},
  \bibinfo{author}{\bibfnamefont{B.~F.} \bibnamefont{Madore}},
  \bibinfo{author}{\bibfnamefont{D.}~\bibnamefont{Hatt}},
  \bibinfo{author}{\bibfnamefont{T.~J.} \bibnamefont{Hoyt}},
  \bibinfo{author}{\bibfnamefont{I.~S.} \bibnamefont{Jang}},
  \bibinfo{author}{\bibfnamefont{R.~L.} \bibnamefont{Beaton}},
  \bibinfo{author}{\bibfnamefont{C.~R.} \bibnamefont{Burns}},
  \bibinfo{author}{\bibfnamefont{M.~G.} \bibnamefont{Lee}},
  \bibinfo{author}{\bibfnamefont{A.~J.} \bibnamefont{Monson}},
  \bibinfo{author}{\bibfnamefont{J.~R.} \bibnamefont{Neeley}},
  \bibnamefont{et~al.}, \bibinfo{journal}{The Astrophysical Journal}
  \textbf{\bibinfo{volume}{882}}, \bibinfo{pages}{34} (\bibinfo{year}{2019}),
  \urlprefix\url{https://doi.org/10.3847/1538-4357/ab2f73}.

\bibitem[{\citenamefont{de~Jaeger et~al.}(2020)\citenamefont{de~Jaeger, Stahl,
  Zheng, Filippenko, Riess, and Galbany}}]{de_Jaeger_2020}
\bibinfo{author}{\bibfnamefont{T.}~\bibnamefont{de~Jaeger}},
  \bibinfo{author}{\bibfnamefont{B.~E.} \bibnamefont{Stahl}},
  \bibinfo{author}{\bibfnamefont{W.}~\bibnamefont{Zheng}},
  \bibinfo{author}{\bibfnamefont{A.~V.} \bibnamefont{Filippenko}},
  \bibinfo{author}{\bibfnamefont{A.~G.} \bibnamefont{Riess}}, \bibnamefont{and}
  \bibinfo{author}{\bibfnamefont{L.}~\bibnamefont{Galbany}},
  \bibinfo{journal}{Monthly Notices of the Royal Astronomical Society}
  \textbf{\bibinfo{volume}{496}}, \bibinfo{pages}{3402} (\bibinfo{year}{2020}),
  \urlprefix\url{https://doi.org/10.1093/mnras/staa1801}.

\bibitem[{\citenamefont{Huang et~al.}(2020)\citenamefont{Huang, Riess, Yuan,
  Macri, Zakamska, Casertano, Whitelock, Hoffmann, Filippenko, and
  Scolnic}}]{Huang_2020}
\bibinfo{author}{\bibfnamefont{C.~D.} \bibnamefont{Huang}},
  \bibinfo{author}{\bibfnamefont{A.~G.} \bibnamefont{Riess}},
  \bibinfo{author}{\bibfnamefont{W.}~\bibnamefont{Yuan}},
  \bibinfo{author}{\bibfnamefont{L.~M.} \bibnamefont{Macri}},
  \bibinfo{author}{\bibfnamefont{N.~L.} \bibnamefont{Zakamska}},
  \bibinfo{author}{\bibfnamefont{S.}~\bibnamefont{Casertano}},
  \bibinfo{author}{\bibfnamefont{P.~A.} \bibnamefont{Whitelock}},
  \bibinfo{author}{\bibfnamefont{S.~L.} \bibnamefont{Hoffmann}},
  \bibinfo{author}{\bibfnamefont{A.~V.} \bibnamefont{Filippenko}},
  \bibnamefont{and} \bibinfo{author}{\bibfnamefont{D.}~\bibnamefont{Scolnic}},
  \bibinfo{journal}{The Astrophysical Journal} \textbf{\bibinfo{volume}{889}},
  \bibinfo{pages}{5} (\bibinfo{year}{2020}),
  \urlprefix\url{https://doi.org/10.3847/1538-4357/ab5dbd}.

\bibitem[{\citenamefont{Kourkchi et~al.}(2020)\citenamefont{Kourkchi, Tully,
  Anand, Courtois, Dupuy, Neill, Rizzi, and Seibert}}]{Kourkchi_2020}
\bibinfo{author}{\bibfnamefont{E.}~\bibnamefont{Kourkchi}},
  \bibinfo{author}{\bibfnamefont{R.~B.} \bibnamefont{Tully}},
  \bibinfo{author}{\bibfnamefont{G.~S.} \bibnamefont{Anand}},
  \bibinfo{author}{\bibfnamefont{H.~M.} \bibnamefont{Courtois}},
  \bibinfo{author}{\bibfnamefont{A.}~\bibnamefont{Dupuy}},
  \bibinfo{author}{\bibfnamefont{J.~D.} \bibnamefont{Neill}},
  \bibinfo{author}{\bibfnamefont{L.}~\bibnamefont{Rizzi}}, \bibnamefont{and}
  \bibinfo{author}{\bibfnamefont{M.}~\bibnamefont{Seibert}},
  \bibinfo{journal}{The Astrophysical Journal} \textbf{\bibinfo{volume}{896}},
  \bibinfo{pages}{3} (\bibinfo{year}{2020}),
  \urlprefix\url{https://doi.org/10.3847/1538-4357/ab901c}.

\bibitem[{\citenamefont{Blakeslee et~al.}(2021)\citenamefont{Blakeslee, Jensen,
  Ma, Milne, and Greene}}]{Blakeslee_2021}
\bibinfo{author}{\bibfnamefont{J.~P.} \bibnamefont{Blakeslee}},
  \bibinfo{author}{\bibfnamefont{J.~B.} \bibnamefont{Jensen}},
  \bibinfo{author}{\bibfnamefont{C.-P.} \bibnamefont{Ma}},
  \bibinfo{author}{\bibfnamefont{P.~A.} \bibnamefont{Milne}}, \bibnamefont{and}
  \bibinfo{author}{\bibfnamefont{J.~E.} \bibnamefont{Greene}},
  \bibinfo{journal}{The Astrophysical Journal} \textbf{\bibinfo{volume}{911}},
  \bibinfo{pages}{65} (\bibinfo{year}{2021}),
  \urlprefix\url{https://doi.org/10.3847/1538-4357/abe86a}.

\bibitem[{\citenamefont{Farren et~al.}(2022)\citenamefont{Farren, Philcox, and
  Sherwin}}]{Farren_2022}
\bibinfo{author}{\bibfnamefont{G.~S.} \bibnamefont{Farren}},
  \bibinfo{author}{\bibfnamefont{O.~H.~E.} \bibnamefont{Philcox}},
  \bibnamefont{and} \bibinfo{author}{\bibfnamefont{B.~D.}
  \bibnamefont{Sherwin}}, \bibinfo{journal}{Physical Review D}
  \textbf{\bibinfo{volume}{105}}, \bibinfo{pages}{063503}
  (\bibinfo{year}{2022}),
  \urlprefix\url{https://doi.org/10.1103/physrevd.105.063503}.

\bibitem[{\citenamefont{Reid et~al.}(2019)\citenamefont{Reid, Pesce, and
  Riess}}]{Reid_2019}
\bibinfo{author}{\bibfnamefont{M.~J.} \bibnamefont{Reid}},
  \bibinfo{author}{\bibfnamefont{D.~W.} \bibnamefont{Pesce}}, \bibnamefont{and}
  \bibinfo{author}{\bibfnamefont{A.~G.} \bibnamefont{Riess}},
  \bibinfo{journal}{The Astrophysical Journal} \textbf{\bibinfo{volume}{886}},
  \bibinfo{pages}{L27} (\bibinfo{year}{2019}),
  \urlprefix\url{https://doi.org/10.3847/2041-8213/ab552d}.

\bibitem[{\citenamefont{Aiola et~al.}(2020)\citenamefont{Aiola, Calabrese,
  Maurin, Naess, Schmitt, Abitbol, Addison, Ade, Alonso, Amiri
  et~al.}}]{Aiola_2020}
\bibinfo{author}{\bibfnamefont{S.}~\bibnamefont{Aiola}},
  \bibinfo{author}{\bibfnamefont{E.}~\bibnamefont{Calabrese}},
  \bibinfo{author}{\bibfnamefont{L.}~\bibnamefont{Maurin}},
  \bibinfo{author}{\bibfnamefont{S.}~\bibnamefont{Naess}},
  \bibinfo{author}{\bibfnamefont{B.~L.} \bibnamefont{Schmitt}},
  \bibinfo{author}{\bibfnamefont{M.~H.} \bibnamefont{Abitbol}},
  \bibinfo{author}{\bibfnamefont{G.~E.} \bibnamefont{Addison}},
  \bibinfo{author}{\bibfnamefont{P.~A.~R.} \bibnamefont{Ade}},
  \bibinfo{author}{\bibfnamefont{D.}~\bibnamefont{Alonso}},
  \bibinfo{author}{\bibfnamefont{M.}~\bibnamefont{Amiri}},
  \bibnamefont{et~al.}, \bibinfo{journal}{Journal of Cosmology and
  Astroparticle Physics} \textbf{\bibinfo{volume}{2020}}, \bibinfo{pages}{047}
  (\bibinfo{year}{2020}),
  \urlprefix\url{https://doi.org/10.1088/1475-7516/2020/12/047}.

\bibitem[{\citenamefont{Yang et~al.}(2019)\citenamefont{Yang, Pan, Vagnozzi,
  Valentino, Mota, and Capozziello}}]{Yang_2019}
\bibinfo{author}{\bibfnamefont{W.}~\bibnamefont{Yang}},
  \bibinfo{author}{\bibfnamefont{S.}~\bibnamefont{Pan}},
  \bibinfo{author}{\bibfnamefont{S.}~\bibnamefont{Vagnozzi}},
  \bibinfo{author}{\bibfnamefont{E.~D.} \bibnamefont{Valentino}},
  \bibinfo{author}{\bibfnamefont{D.~F.} \bibnamefont{Mota}}, \bibnamefont{and}
  \bibinfo{author}{\bibfnamefont{S.}~\bibnamefont{Capozziello}},
  \bibinfo{journal}{Journal of Cosmology and Astroparticle Physics}
  \textbf{\bibinfo{volume}{2019}}, \bibinfo{pages}{044} (\bibinfo{year}{2019}),
  \urlprefix\url{https://doi.org/10.1088/1475-7516/2019/11/044}.

\bibitem[{\citenamefont{Gayathri et~al.}(2020)\citenamefont{Gayathri, Healy,
  Lange, O'Brien, Szczepanczyk, Bartos, Campanelli, Klimenko, Lousto, and
  O'Shaughnessy}}]{Gayathri_2020}
\bibinfo{author}{\bibfnamefont{V.}~\bibnamefont{Gayathri}},
  \bibinfo{author}{\bibfnamefont{J.}~\bibnamefont{Healy}},
  \bibinfo{author}{\bibfnamefont{J.}~\bibnamefont{Lange}},
  \bibinfo{author}{\bibfnamefont{B.}~\bibnamefont{O'Brien}},
  \bibinfo{author}{\bibfnamefont{M.}~\bibnamefont{Szczepanczyk}},
  \bibinfo{author}{\bibfnamefont{I.}~\bibnamefont{Bartos}},
  \bibinfo{author}{\bibfnamefont{M.}~\bibnamefont{Campanelli}},
  \bibinfo{author}{\bibfnamefont{S.}~\bibnamefont{Klimenko}},
  \bibinfo{author}{\bibfnamefont{C.}~\bibnamefont{Lousto}}, \bibnamefont{and}
  \bibinfo{author}{\bibfnamefont{R.}~\bibnamefont{O'Shaughnessy}}
  (\bibinfo{year}{2020}), \urlprefix\url{https://arxiv.org/abs/2009.14247}.

\bibitem[{\citenamefont{Wong et~al.}(2019)\citenamefont{Wong, Suyu, Chen, Rusu,
  Millon, Sluse, Bonvin, Fassnacht, Taubenberger, Auger et~al.}}]{Wong_2019}
\bibinfo{author}{\bibfnamefont{K.~C.} \bibnamefont{Wong}},
  \bibinfo{author}{\bibfnamefont{S.~H.} \bibnamefont{Suyu}},
  \bibinfo{author}{\bibfnamefont{G.~C.-F.} \bibnamefont{Chen}},
  \bibinfo{author}{\bibfnamefont{C.~E.} \bibnamefont{Rusu}},
  \bibinfo{author}{\bibfnamefont{M.}~\bibnamefont{Millon}},
  \bibinfo{author}{\bibfnamefont{D.}~\bibnamefont{Sluse}},
  \bibinfo{author}{\bibfnamefont{V.}~\bibnamefont{Bonvin}},
  \bibinfo{author}{\bibfnamefont{C.~D.} \bibnamefont{Fassnacht}},
  \bibinfo{author}{\bibfnamefont{S.}~\bibnamefont{Taubenberger}},
  \bibinfo{author}{\bibfnamefont{M.~W.} \bibnamefont{Auger}},
  \bibnamefont{et~al.}, \bibinfo{journal}{Monthly Notices of the Royal
  Astronomical Society} \textbf{\bibinfo{volume}{498}}, \bibinfo{pages}{1420}
  (\bibinfo{year}{2019}),
  \urlprefix\url{https://doi.org/10.1093/mnras/stz3094}.

\bibitem[{\citenamefont{Scolnic et~al.}(2018)\citenamefont{Scolnic, Jones,
  Rest, Pan, Chornock, Foley, Huber, Kessler, Narayan, Riess
  et~al.}}]{scolnic-2018}
\bibinfo{author}{\bibfnamefont{D.~M.} \bibnamefont{Scolnic}},
  \bibinfo{author}{\bibfnamefont{D.~O.} \bibnamefont{Jones}},
  \bibinfo{author}{\bibfnamefont{A.}~\bibnamefont{Rest}},
  \bibinfo{author}{\bibfnamefont{Y.~C.} \bibnamefont{Pan}},
  \bibinfo{author}{\bibfnamefont{R.}~\bibnamefont{Chornock}},
  \bibinfo{author}{\bibfnamefont{R.~J.} \bibnamefont{Foley}},
  \bibinfo{author}{\bibfnamefont{M.~E.} \bibnamefont{Huber}},
  \bibinfo{author}{\bibfnamefont{R.}~\bibnamefont{Kessler}},
  \bibinfo{author}{\bibfnamefont{G.}~\bibnamefont{Narayan}},
  \bibinfo{author}{\bibfnamefont{A.~G.} \bibnamefont{Riess}},
  \bibnamefont{et~al.}, \bibinfo{journal}{Astrophys. J.}
  \textbf{\bibinfo{volume}{859}}, \bibinfo{pages}{101} (\bibinfo{year}{2018}),
  \urlprefix\url{https://iopscience.iop.org/article/10.3847/1538-4357/aab9bb}.

\bibitem[{\citenamefont{Scolnic et~al.}(2022)\citenamefont{Scolnic, Brout,
  Carr, Riess, Davis, Dwomoh, Jones, Ali, Charvu, Chen et~al.}}]{Scolnic_2022}
\bibinfo{author}{\bibfnamefont{D.}~\bibnamefont{Scolnic}},
  \bibinfo{author}{\bibfnamefont{D.}~\bibnamefont{Brout}},
  \bibinfo{author}{\bibfnamefont{A.}~\bibnamefont{Carr}},
  \bibinfo{author}{\bibfnamefont{A.~G.} \bibnamefont{Riess}},
  \bibinfo{author}{\bibfnamefont{T.~M.} \bibnamefont{Davis}},
  \bibinfo{author}{\bibfnamefont{A.}~\bibnamefont{Dwomoh}},
  \bibinfo{author}{\bibfnamefont{D.~O.} \bibnamefont{Jones}},
  \bibinfo{author}{\bibfnamefont{N.}~\bibnamefont{Ali}},
  \bibinfo{author}{\bibfnamefont{P.}~\bibnamefont{Charvu}},
  \bibinfo{author}{\bibfnamefont{R.}~\bibnamefont{Chen}}, \bibnamefont{et~al.},
  \bibinfo{journal}{The Astrophysical Journal} \textbf{\bibinfo{volume}{938}},
  \bibinfo{pages}{113} (\bibinfo{year}{2022}),
  \urlprefix\url{https://doi.org/10.3847/1538-4357/ac8b7a}.

\bibitem[{\citenamefont{Dainotti
  et~al.}(2022{\natexlab{b}})\citenamefont{Dainotti, Sarracino, and
  Capozziello}}]{Dainotti_2022e}
\bibinfo{author}{\bibfnamefont{M.~G.} \bibnamefont{Dainotti}},
  \bibinfo{author}{\bibfnamefont{G.}~\bibnamefont{Sarracino}},
  \bibnamefont{and}
  \bibinfo{author}{\bibfnamefont{S.}~\bibnamefont{Capozziello}},
  \bibinfo{journal}{Publications of the Astronomical Society of Japan}
  \textbf{\bibinfo{volume}{74}}, \bibinfo{pages}{1095}
  (\bibinfo{year}{2022}{\natexlab{b}}),
  \urlprefix\url{https://doi.org/10.1093/pasj/psac057}.

\bibitem[{\citenamefont{Dainotti
  et~al.}(2022{\natexlab{c}})\citenamefont{Dainotti, Lenart, Chraya, Sarracino,
  Nagataki, Fraija, Capozziello, and Bogdan}}]{Dainotti_2022f}
\bibinfo{author}{\bibfnamefont{M.~G.} \bibnamefont{Dainotti}},
  \bibinfo{author}{\bibfnamefont{A.~L.} \bibnamefont{Lenart}},
  \bibinfo{author}{\bibfnamefont{A.}~\bibnamefont{Chraya}},
  \bibinfo{author}{\bibfnamefont{G.}~\bibnamefont{Sarracino}},
  \bibinfo{author}{\bibfnamefont{S.}~\bibnamefont{Nagataki}},
  \bibinfo{author}{\bibfnamefont{N.}~\bibnamefont{Fraija}},
  \bibinfo{author}{\bibfnamefont{S.}~\bibnamefont{Capozziello}},
  \bibnamefont{and} \bibinfo{author}{\bibfnamefont{M.}~\bibnamefont{Bogdan}},
  \bibinfo{journal}{Monthly Notices of the Royal Astronomical Society}
  \textbf{\bibinfo{volume}{518}}, \bibinfo{pages}{2201}
  (\bibinfo{year}{2022}{\natexlab{c}}),
  \urlprefix\url{https://doi.org/10.1093/mnras/stac2752}.

\bibitem[{\citenamefont{Bargiacchi et~al.}(2022)\citenamefont{Bargiacchi,
  Benetti, Capozziello, Lusso, Risaliti, and Signorini}}]{Bargiacchi_2022}
\bibinfo{author}{\bibfnamefont{G.}~\bibnamefont{Bargiacchi}},
  \bibinfo{author}{\bibfnamefont{M.}~\bibnamefont{Benetti}},
  \bibinfo{author}{\bibfnamefont{S.}~\bibnamefont{Capozziello}},
  \bibinfo{author}{\bibfnamefont{E.}~\bibnamefont{Lusso}},
  \bibinfo{author}{\bibfnamefont{G.}~\bibnamefont{Risaliti}}, \bibnamefont{and}
  \bibinfo{author}{\bibfnamefont{M.}~\bibnamefont{Signorini}},
  \bibinfo{journal}{Monthly Notices of the Royal Astronomical Society}
  \textbf{\bibinfo{volume}{515}}, \bibinfo{pages}{1795–1806}
  (\bibinfo{year}{2022}),
  \urlprefix\url{https://doi.org/10.1093/mnras/stac1941}.

\bibitem[{\citenamefont{Dainotti
  et~al.}(2022{\natexlab{d}})\citenamefont{Dainotti, Nielson, Sarracino,
  Rinaldi, Nagataki, Capozziello, Gnedin, and Bargiacchi}}]{Dainotti_2022c}
\bibinfo{author}{\bibfnamefont{M.~G.} \bibnamefont{Dainotti}},
  \bibinfo{author}{\bibfnamefont{V.}~\bibnamefont{Nielson}},
  \bibinfo{author}{\bibfnamefont{G.}~\bibnamefont{Sarracino}},
  \bibinfo{author}{\bibfnamefont{E.}~\bibnamefont{Rinaldi}},
  \bibinfo{author}{\bibfnamefont{S.}~\bibnamefont{Nagataki}},
  \bibinfo{author}{\bibfnamefont{S.}~\bibnamefont{Capozziello}},
  \bibinfo{author}{\bibfnamefont{O.~Y.} \bibnamefont{Gnedin}},
  \bibnamefont{and}
  \bibinfo{author}{\bibfnamefont{G.}~\bibnamefont{Bargiacchi}},
  \bibinfo{journal}{Monthly Notices of the Royal Astronomical Society}
  \textbf{\bibinfo{volume}{514}}, \bibinfo{pages}{1828}
  (\bibinfo{year}{2022}{\natexlab{d}}),
  \urlprefix\url{https://doi.org/10.1093/mnras/stac1141}.

\bibitem[{\citenamefont{Califano
  et~al.}(2022{\natexlab{a}})\citenamefont{Califano, de~Martino, Vernieri, and
  Capozziello}}]{Califano1}
\bibinfo{author}{\bibfnamefont{M.}~\bibnamefont{Califano}},
  \bibinfo{author}{\bibfnamefont{I.}~\bibnamefont{de~Martino}},
  \bibinfo{author}{\bibfnamefont{D.}~\bibnamefont{Vernieri}}, \bibnamefont{and}
  \bibinfo{author}{\bibfnamefont{S.}~\bibnamefont{Capozziello}}
  (\bibinfo{year}{2022}{\natexlab{a}}),
  \urlprefix\url{https://arxiv.org/abs/2208.13999}.

\bibitem[{\citenamefont{Califano
  et~al.}(2022{\natexlab{b}})\citenamefont{Califano, de~Martino, Vernieri, and
  Capozziello}}]{Califano2}
\bibinfo{author}{\bibfnamefont{M.}~\bibnamefont{Califano}},
  \bibinfo{author}{\bibfnamefont{I.}~\bibnamefont{de~Martino}},
  \bibinfo{author}{\bibfnamefont{D.}~\bibnamefont{Vernieri}}, \bibnamefont{and}
  \bibinfo{author}{\bibfnamefont{S.}~\bibnamefont{Capozziello}},
  \bibinfo{journal}{Monthly Notices of the Royal Astronomical Society}
  \textbf{\bibinfo{volume}{518}}, \bibinfo{pages}{3372}
  (\bibinfo{year}{2022}{\natexlab{b}}),
  \urlprefix\url{https://doi.org/10.10933/mnras/stac3230}.

\bibitem[{\citenamefont{Lenart et~al.}(2023)\citenamefont{Lenart, Bargiacchi,
  Dainotti, Nagataki, and Capozziello}}]{Bargiacchi_3}
\bibinfo{author}{\bibfnamefont{A.~L.} \bibnamefont{Lenart}},
  \bibinfo{author}{\bibfnamefont{G.}~\bibnamefont{Bargiacchi}},
  \bibinfo{author}{\bibfnamefont{M.~G.} \bibnamefont{Dainotti}},
  \bibinfo{author}{\bibfnamefont{S.}~\bibnamefont{Nagataki}}, \bibnamefont{and}
  \bibinfo{author}{\bibfnamefont{S.}~\bibnamefont{Capozziello}},
  \bibinfo{journal}{Astrophysical Journal Supplement Series}
  \textbf{\bibinfo{volume}{264}}, \bibinfo{pages}{46} (\bibinfo{year}{2023}),
  \urlprefix\url{https://doi.org/10.3847/1538-4365/aca404}.

\bibitem[{\citenamefont{Di$\thinspace$Valentino}(2021)}]{Di_Valentino_2021}
\bibinfo{author}{\bibfnamefont{E.}~\bibnamefont{Di$\thinspace$Valentino}},
  \bibinfo{journal}{Monthly Notices of the Royal Astronomical Society}
  \textbf{\bibinfo{volume}{502}}, \bibinfo{pages}{2065} (\bibinfo{year}{2021}),
  \urlprefix\url{https://doi.org/10.1093/mnras/stab187}.

\end{thebibliography}

\end{document}